%% file: cosine-cosmogenics.tex
\documentclass[preprint,5p,twocolumn]{elsarticle}

\usepackage{amssymb}

\usepackage[utf8]{inputenc}
\usepackage{stfloats}
\usepackage{lipsum}
\usepackage{amsmath}
\usepackage{xcolor}

\journal{Astroparticle Physics}

\begin{document}

\begin{frontmatter}



\title{Study of cosmogenic radionuclides in the  COSINE-100 NaI(Tl) detectors}

%
\input{authors_May2019}

\begin{abstract}
 COSINE-100 is a direct detection dark matter search experiment that uses a 106~kg array of eight NaI(Tl) crystals
 that are kept underground at the Yangyang Underground Laboratory to avoid cosmogenic activation
 of radioisotopes by cosmic rays. Even though the cosmogenic activity is declining with
 time, there are still significant background rates from the remnant nuclides.  
 In this paper, we report measurements of cosmogenic isotope contaminations with less than one year
 half-lives that are based on extrapolations
 of the time dependent activities of their characteristic energy peaks to activity rates at the time the crystals
 were deployed underground. For longer-lived $^{109}$Cd ($T_{1/2}=1.27$~y) and
 $^{22}$Na ($T_{1/2}=2.6$~y), we investigate time correlations and coincidence events due to several emissions. 
 The inferred sea-level production rates are compared with
 calculations based on the ACTIVIA and MENDL-2 model calculations and experimental data. 
 The results from different approaches are in reasonable agreement with each other.
 For $^{3}$H, which has a long, 12.3~year half-life, we evaluated the activity levels 
 and the exposure times that are in reasonable agreement with the time period estimated for each crystal's exposure.
\end{abstract}

\begin{keyword}
Cosmogenic radionuclide, activity, production rate, COSINE-100


 
\end{keyword}

\end{frontmatter}


\section{Introduction}
\label{sec:intro}
\begin{figure*}[ht]
\begin{center}
\begin{tabular}{cc}
\includegraphics[width=0.45\textwidth]{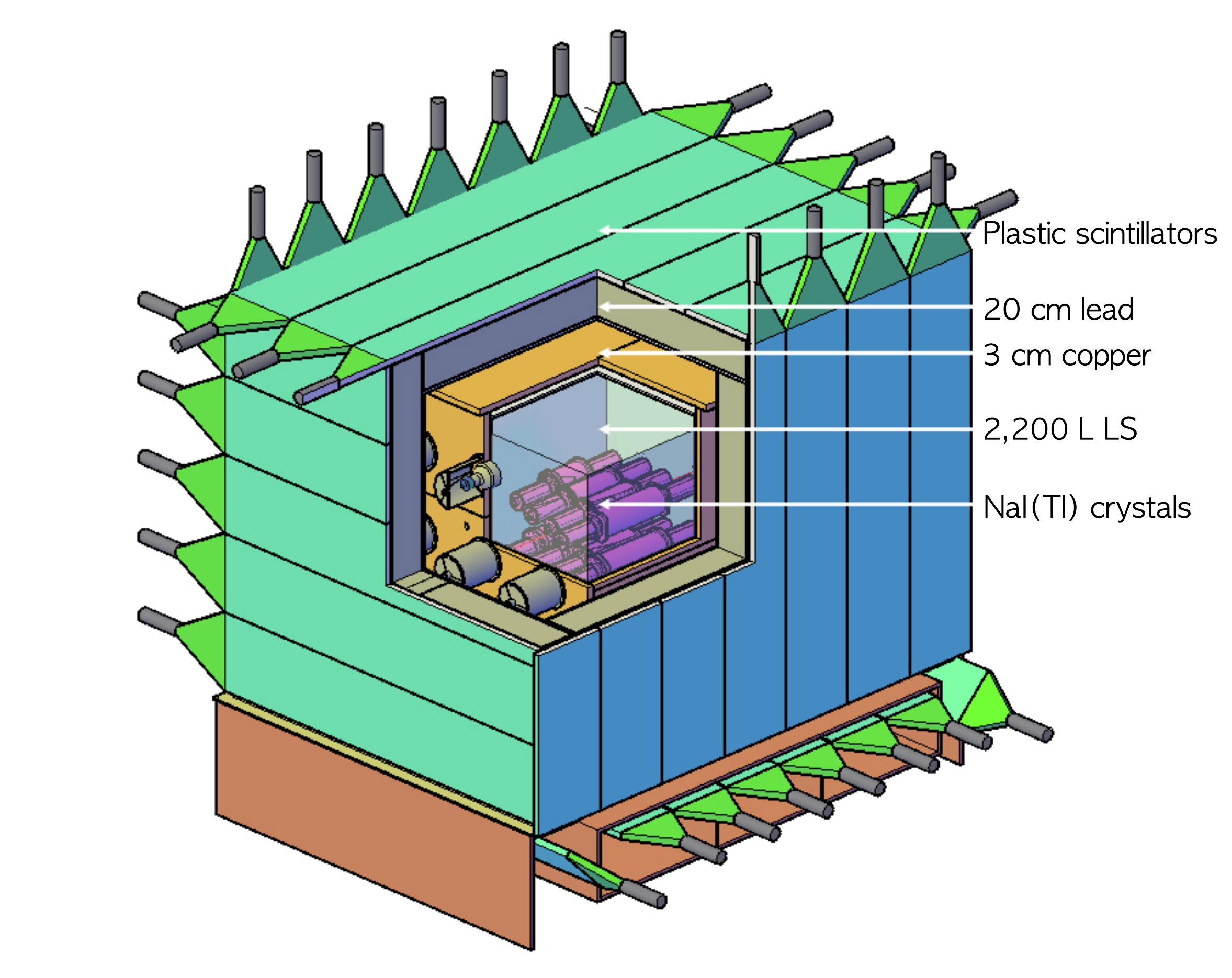} &
\includegraphics[width=0.45\textwidth]{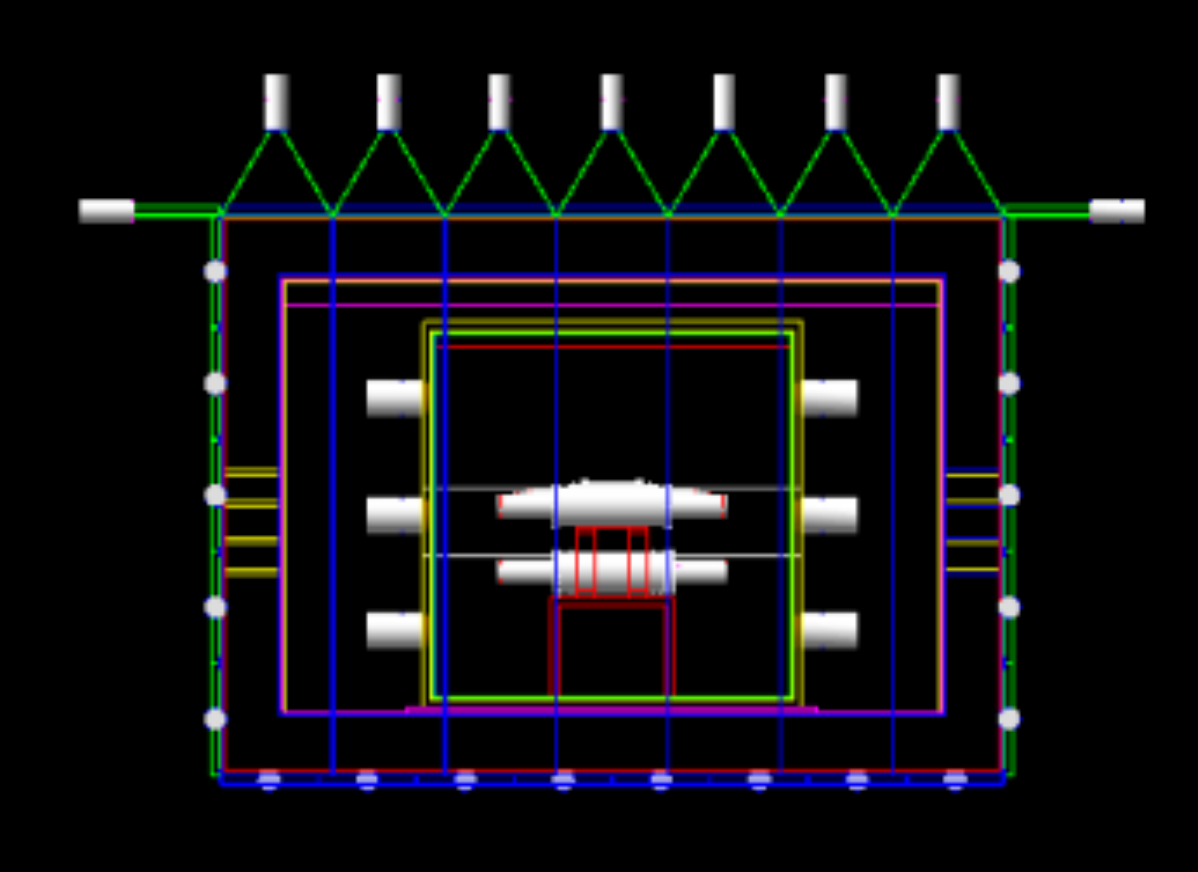} \\
(a)  & (b)  \\
\end{tabular}
\end{center}
\caption{
  (a) The COSINE-100 detector.
  From outside inward, the four shielding layers include: 
  3~cm thick plastic scintillator panels (green), 20~cm of lead (khaki), a 3~cm thick copper box (orange), and
  liquid scintillator (light blue). (b) A side view of the detector geometry used in the Geant4 simulations.
}
\label{detector-setup}
\end{figure*}

There are a number of experiments that search for direct evidence for dark matter particles in the halo of our Galaxy
by looking for nuclei recoiling from dark matter–nucleus scattering~\cite{jungman96, gaitskell04} and report null results.
One notable exception is the DAMA/LIBRA experiment that has consistently reported the observation of an
annual event-rate modulation, that could be interpreted as dark-matter signal, in an array of NaI(Tl) crystal
detectors with a statistical significance that is now more than 12.9~$\sigma$~\cite{Bernabei:2013xsa,Bernabei:2018yyw}.
Although this signal has persisted for over two decades and for three different configurations of the detector,
it remains controversial because it is in conflict with the bounds from other direct detection experiments
using different target materials~\cite{sckim12,PhysRevLett.118.021303,Cui:2017,Aprile:2018,DarkSide50:2018,SuperCDMS:2018} and indirect
searches~\cite{Choi:2015ara}. However, since these conflicts depend on the details of the models for dark matter-nucleus scattering~\cite{Baum:2018ekm}
and the properties of the galactic dark matter halo~\cite{PhysRevD.33.3495,Savage:2006qr,Freese:2012xd},  
a conclusive statement about the DAMA/LIBRA signal can only be made by conducting an independent experiment using the
same NaI(Tl) target material.  This is the prime motivation of COSINE-100 and a number of other NaI(Tl)-crystal-based
experiments~\cite{deSouza:2016fxg,Coarasa:2019,sabre:2019,Fushimi:2015sew,Angloher:2016}

COSINE-100 is a dark matter direct detection experiment~\cite{Adhikari:2017esn,Adhikari:2018ljm} that uses a 106~kg array
of eight low-background  NaI(Tl) crystals situated in a 2000 liter liquid scintillator veto counter. The experiment is
located 700~m underground at the Yangyang Underground Laboratory (Y2L), where it has been operating since September 2016.
The search for an annual modulation signal requires a complete understanding of background sources and their time dependence.
To accomplish this, a complete simulation that accurately models the background energy spectra measured
in the detector is required~\cite{cosinebg}. In addition to backgrounds from long-lived radioactive contaminations in the
crystal bulk and surfaces, we have to deal with time-dependent backgrounds from short-lived cosmogenically activated
radionuclides. These are isotopes that are created by interactions of cosmic rays with stable nuclides in the detector material.
In  COSINE-100, almost all of the cosmogenic isotopes come from cosmic ray interactions with either Na or I nuclei.

This paper is organized as follows.
The COSINE-100 detector is described in section~\ref{sec:2}.
In section~\ref{sec:3}, the cosmogenic isotopes that are produced in NaI(Tl) are listed and
the determination of the activity levels at the time of their initial deployment underground at Y2L is described.
The use of these initial activity levels to infer
production rates for cosmogenic isotopes at sea level and 
their comparison with ACTIVIA and MENDL-2 calculations~\cite{activia,mendl-2} and with experimental data are discussed in
section~\ref{sec:4}.
The fitted activities of $^{3}$H and $^{129}$I from the background modeling are evaluated
in section~\ref{sec:5} and conclusions are provided in section~\ref{sec:conc}.

\section{The COSINE-100 experimental setup}
\label{sec:2}

\begin{table*}
\begin{center}
\caption{
Mass, dimensions, powder, surface exposure, and underground radioactivity cooling times for each one of the analyzed crystals (see text).
}
\label{crystal}
\begin{tabular}{c|c|c|c|c|c}
\hline 
Crystal & Mass & Size & Powder type & Exposure time & Radioactivity \\ 
&  (kg) & (diameter$\times$length)  & & (years)
& cooling time at Y2L \\
& & (inches) & & & (years) \\ \hline
Crystal-1 & 8.3 & 5.0$\times$7.0 & AS-B & 2.17 & 3 \\
Crystal-2 & 9.2 & 4.2$\times$11.0 & AS-C & 0.92 & 2.75 \\
Crystal-3 & 9.2 & 4.2$\times$11.0 & AS-WSII & $>$ 0.92 & 1.2 \\
Crystal-4 & 18.0 & 5.0$\times$15.3 & AS-WSII & 1.83 & 0.5 \\
Crystal-6 & 12.5 & 4.8$\times$11.8 & AS-WSIII & 0.5 & 0.6 \\
Crystal-7 & 12.5 & 4.8$\times$11.8 & AS-WSIII & 0.5 & 0.6 \\ \hline
\end{tabular}
\end{center}
\end{table*}

The experimental setup of COSINE-100, shown in Fig.~\ref{detector-setup}(a), is described in detail in
Ref.~\cite{Adhikari:2017esn}.  Eight NaI(Tl) crystals, arranged in two layers, are located in the
middle of a four-layer shielding structure. From outside inward, this comprises plastic scintillator
panels, a lead-brick castle, a copper box, and a tank of scintillating liquid. The eight encapsulated NaI(Tl) crystal
assemblies and their support table are immersed in the scintillating liquid that serves both as an active
veto and a passive shield. The eight NaI(Tl) crystals were grown from powder provided by Alpha
Spectra~(AS). Two crystals~(Crystal-5 and Crystal-8) are not considered in this paper because their
low light yields result in poorer energy resolution and 
because of their relatively high background contamination levels, especially at low energies.

Since the detailed cosmic ray exposure history of each crystal is unknown, we estimated the time period for each crystal's
exposure, listed in Table~\ref{crystal}, from the time between the powder production by Alpha Spectra
at Grand Junction, Colorado, to the date delivered to Y2L. 
We considered that the preparation of the NaI powder precedes the crystal growth date by 2 months.
It includes 30 days as the duration of transportation to Y2L. 
Since Crystal-3 has a complicated exposure history, having been repaired once before deployment at Y2L, 
we can only be certain that the corresponding period is more than 9 months.
The radioactivity cooling time for each crystal between delivery to Y2L and the start of data-taking
is also listed in Table~\ref{crystal}.   

The six crystals that are considered in this analysis have light yields of about
15~photoelectrons/keV; the energy threshold for an accepted signal from a crystal is 2~keV.
Selection criteria that are used to isolate scintillation-light generated
signals from photomultiplier tube noise are described in detail in Ref.~\cite{Adhikari:2017esn}. 
Events that have above-threshold signals in only one of the crystals
and none in any of the other crystals or the liquid scintillator are
classified as {\it single-hit} events. 
Those with above-threshold signals in more than one crystal and/or the liquid
scintillator are classified as {\it multiple-hit} events.

Monte Carlo simulations based on the Geant4 toolkit~\cite{Agostinelli:2002hh} are used
to better understand the background spectra from the cosmogenic isotopes in the crystals;
the geometry used for these simulations is shown in Fig.~\ref{detector-setup}(b).

\section{Cosmogenic radionuclides and initial activities}
\label{sec:3}
\begin{table*}
\begin{center}
\caption{
Cosmogenic radionuclides in the NaI(Tl) crystals identified in other studies and taken into consideration here.
}
\label{cosmogenic}
\begin{tabular}{c|c|c}
\hline 
Cosmogenic & Half-life~\cite{DDEP, Ohya:2010, LNDS} & Decay type \\ 
isotopes & (days) & \& Emissions energy \\ \hline
$^{125}$I & 59.4 & EC, 35.5+31.7=67.2~keV  \\
$^{121}$Te & 19.17 & EC, 4.1--4.7 and 30.5~keV  \\
$^{121m}$Te & 164.2 & EC, 4.1--4.7 and 30.5~keV \\
$^{123m}$Te & 119.3 & IT, 247~keV\\
$^{125m}$Te & 57.4 & IT, 145~keV \\
$^{127m}$Te & 106.1 & IT, 88~keV \\ 
$^{113}$Sn & 115.1 & EC, 3.7--4.2 and 28~keV \\ \hline
$^{109}$Cd & 462 & EC, 25.5 and 88~keV\\
$^{22}$Na & 950 & $\beta^{+}$, 511 and 1274.6~keV \\
$^{3}$H & 4494 & $\beta^{-}$ \\
$^{129}$I & 1.57$\times$10$^{7}$ yr & $\beta^{-}$ \\ \hline
\end{tabular}
\end{center}
\end{table*}

\begin{figure*}[ht]
\begin{center}
\includegraphics[width=1.0\textwidth]{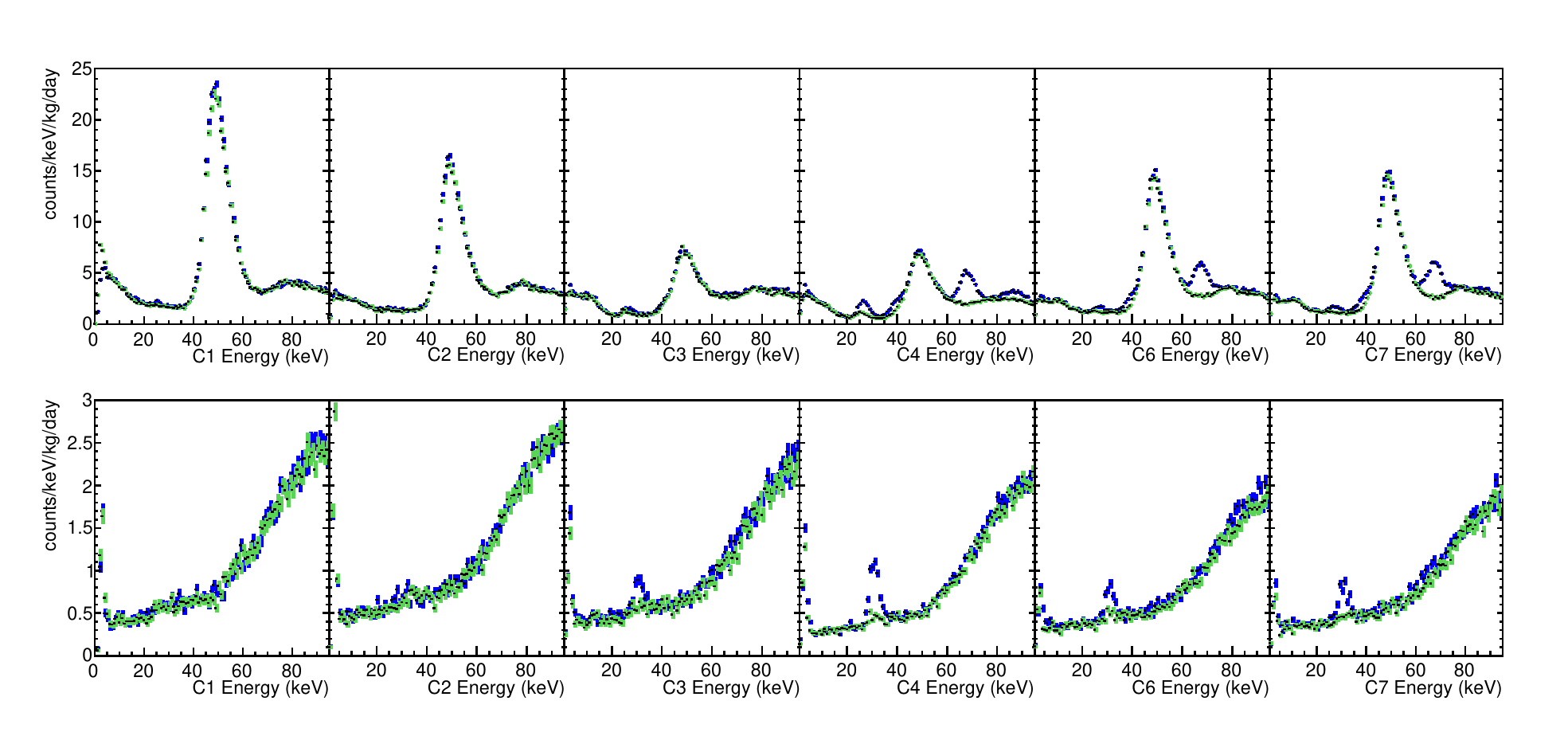}
\caption{
  Background spectra for six NaI(Tl) crystals during the first (blue points) and the last (green points) 25 days of the dataset
  taken from October 21, 2016 to July 18, 2018. The upper plots show single-hit events and the lower
  ones show multiple-hit events.}
\label{fig:bkgComp}
\end{center}
\end{figure*}

\begin{figure}[!b] 
\begin{center}
\includegraphics[width=0.5\textwidth]{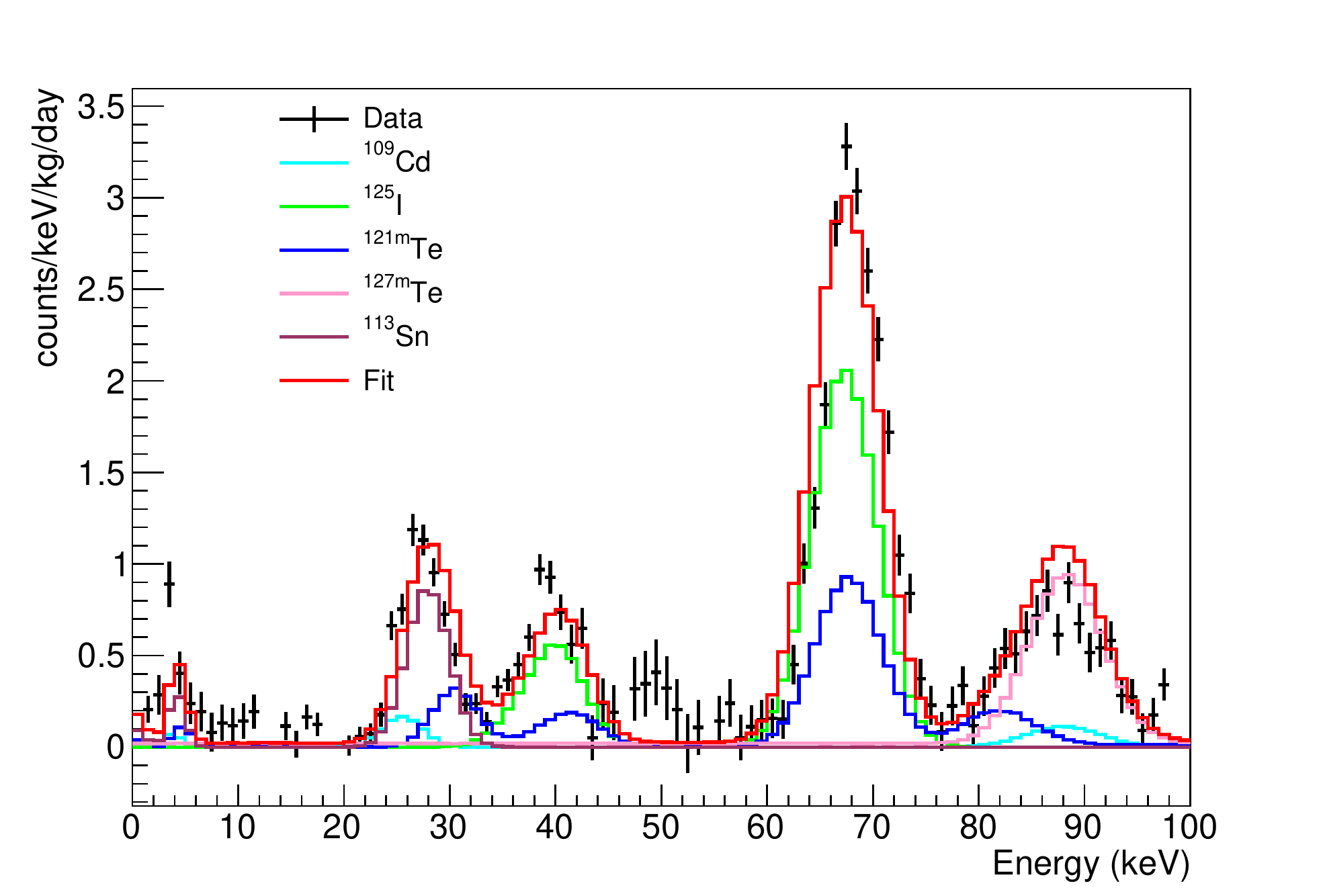}
\caption{
Difference between the first and last 25~day spectra of the 1.7~year data for Crystal-4 (in black) together with the fitted spectrum (in red) from several simulated individual contributions (in several colors).
The subtracted spectra in Crystals-4, 6, and 7 are similar to each other due to their relatively short cooling times underground. 
However, the short-lived isotopes are not expected to contribute to either Crystal-1 or Crystal-2 due to their relatively long cooling times underground. 
}
\label{fig:cosmogenics}
\end{center}
\end{figure}

Although the eight NaI(Tl) crystals had underground radioactivity cooling times that range from several months to
three years, there are still backgrounds from the long-lived cosmogenic isotopes that were activated
by cosmic rays while they were on the surface.

To understand these backgrounds, we first considered the list of cosmogenic radioactive isotopes that are produced in
NaI(Tl) reported in Ref.~\cite{walter-thesis,cosmogenic-amare15,cosmogenic-villar18,cosmogenic-amare18}.
In Table~\ref{cosmogenic}, we list the contributing cosmogenic isotopes with their half lives and decay modes; short-lived isotopes,
for which half lives are less than a year, are $^{125}$I, $^{121}$Te, $^{121m}$Te, $^{123m}$Te, $^{125m}$Te, $^{127m}$Te, and
$^{113}$Sn and long-lived isotopes are $^{109}$Cd, $^{22}$Na, $^{3}$H, and $^{129}$I. 
Since there are no characteristic peaks from the decay of $^{123m}$Te/$^{125m}$Te in the low energy below 100 keV 
their contributions are negligible in all crystals in Table~\ref{crystal} and, thus, they are not further considered in the analysis. 

The short-lived ($T_{1/2}<1$~y) isotopes are not expected to contribute significantly
to either Crystal-1 or Crystal-2 because their cooling times
are long enough to reduce these activities to a
negligible level. However, we expect some backgrounds from the short-lived isotopes in other crystals because
their production rates at sea level, 
as estimated in~\cite{walter-thesis,cosmogenic-amare15,cosmogenic-villar18,cosmogenic-amare18}, 
are high and their cooling times are less than or equal to a year. 

Data points in Fig.~\ref{fig:bkgComp} show the energy spectra for the six considered NaI(Tl) crystals
during the first (blue) and last (green) 25~day segments of the dataset taken between October 21, 2016
and July 18, 2018.  A significant reduction of peaks from short-lived cosmogenic isotopes in
Crystals~4,~6, and~7 for both single- and multiple-hit events is evident, while the differences for
Crystals~1 and 2 are small, as expected.   To associate the specific peaks with its cosmogenic nuclide,
we simulated each isotope in Table~\ref{cosmogenic} as a radioactive contaminant randomly distributed inside
the NaI(Tl) crystal bulk. Fig.~\ref{fig:cosmogenics} shows the differences between the 
initial and final data segments for Crystal-4.  
The subtracted spectrum is well fitted by the simulated
cosmogenic components that are treated as parameters floating in the fit, thereby validating our selection of the main cosmogenic contributors
to the low-energy  single-hit distribution. 
However, 
the derived weight of each isotope from the fit are not further considered in the analysis.
Those two structures at about 12 and 48 keV in Fig.~\ref{fig:cosmogenics} are characteristic of $^{210}$Pb. It decays over time with the half-life of 22.3~yr and 
there is a little difference between the first and last 25 day spectra of the 1.7 years data.

Four long-lived nuclides, $^{109}$Cd, $^{22}$Na, $^{3}$H, and $^{129}$I have low energy deposits and are, therefore, potentially troublesome.  
It is essential to understand their background contributions to the low energy spectra regions, 
especially in the (2--6)~keV dark matter signal region of interest (ROI). 
The beta-decay spectrum of tritium has an endpoint energy of 18 keV and the electron capture decay of $^{22}$Na produces 0.87~keV emissions. 
The beta decay of $^{129}$I to $^{129}$Xe$^{*}$ is followed by $^{129}$Xe$^{*}$ transitioning to the stable $^{129}$Xe isotope via the emission of a 39.6~keV $\gamma$-ray. 
Its spectral feature from this process has a distribution with a peak around $\sim$45 keV.  
The electron capture decay of  $^{109}$Cd contribute peaks at 25.5 and around 3.5~keV. 

Because it is impossible to compute the initial activities of the cosmogenic  radioisotopes from the production rates
without knowing their detailed cosmic ray exposure conditions: i.e., time, location, altitude, etc.~\cite{walter-thesis},
we, when possible, extrapolated the time-dependent reduction of characteristic peaks from their decays to determine their
activity levels at the time of their initial deployment at Y2L. 
For the activities of the long-lived $^{22}$Na and $^{109}$Cd isotopes we investigated temporal and spatial
correlations of characteristic $\gamma$/X-rays peaks produced in their decays. 
The details of these technique are discussed in the following sections.

\input{analysis1_v1.9.tex}
\input{analysis2_v1.9.tex}

\section{Results and comparisons for production rates}
\label{sec:4}
In section~\ref{sec:3} we describe the determination of the crystals' cosmogenic isotope
activities at the time they were first deployed underground at Y2L. 
However, since we do not know the details of their previous exposure conditions,
such as times, locations, and altitudes, these cannot be directly related to production rates
or saturation activity levels. 
But an attempt to extract sea level production rates has been made from a simplified mathematical model for production and decay of radionuclides.

\begin{table*}[ht]
\begin{center}
    \caption{Production rate R$_{s}$ [kg$^{-1}$ d$^{-1}$] at sea level.
    \label{production_rate}
    }
    \begin{tabular}{c|cccccc}
      \hline            
      &  $^{22}$Na  &  $^{109}$Cd   &  $^{125}$I  &  $^{121m}$Te  &  $^{127m}$Te & $^{113}$Sn  \\ \hline 
      Crystal-1 & 132.0$\pm$24.0 & 1.7$\pm$1.1 & & &   \\
      Crystal-2 & 148.5$\pm$44.9 & 0.6$\pm$1.2 & & &   \\
      Crystal-3 & 114.5$\pm$19.7 & 4.7$\pm$0.6 & 280.1$\pm$29.3 & 31.1$\pm$5.5 & 26.9$\pm$4.8 & 5.1$\pm$1.6 \\
      Crystal-4 & 81.0$\pm$12.7 & 3.7$\pm$0.3 & 104.2$\pm$3.7 & 24.9$\pm$1.6 & 13.5$\pm$0.7 & 4.1$\pm$1.6 \\
      Crystal-6 & 144.0$\pm$31.2 & 1.8$\pm$0.8 & 184.7$\pm$6.3 & 23.5$\pm$3.5 & 16.3$\pm$1.5 & 7.1$\pm$0.5 \\
      Crystal-7 & 151.0$\pm$52.1 & 1.8$\pm$ 0.6& 194.0$\pm$6.3 & 22.3$\pm$3.5 & 15.0$\pm$1.5 & 5.3$\pm$0.5 \\ \hline
      ACTIVIA & 66 & 4.8 & 221 & 93 & 93 & 9 \\
      MENDL-2 & & 4.8 & 208 & 102 &  \\
      ANAIS measurement~\cite{cosmogenic-amare15,cosmogenic-villar18,anais112:2019} & 45.1$\pm$1.9 & 2.38$\pm$0.20 & 220$\pm$10 & 23.5$\pm$0.8 & 10.2$\pm$0.4 & 4.53$\pm$0.40  \\
      DM-Ice17 measurement~\cite{walter-thesis} & & & 230 & 25 & $<$ 9 & 16  \\  \hline        
    \end{tabular}
\end{center}    
\end{table*}

The production rate $R$ for activation of an isotope can be expressed as
\begin{equation}
\label{eq:4.1} 
R \propto \int{\sigma (E) \cdot \Phi (E) \cdot dE} 
\end{equation}
where $\sigma$ is the neutron capture cross section and $\Phi$ is the cosmic-ray neutron flux. 
Since cosmic-ray neutron flux $\Phi$ depends on altitude, location, and time the production rate $R$ at any arbitrary location can be calculated by scaling the reference production rate R$_{s}$ at sea level, 
\begin{equation}
\label{eq:4.1}
R = f \cdot R_{s}
\end{equation}
The produced nuclide then decays according to the standard decay law, and the net rate of change for
the number of existing radioactive nuclei $N$ is by the differential equation
\begin{equation}
\label{eq:4.2}
\frac{dN}{dt} = - \lambda \cdot N + R~, 	
\end{equation}
where $\lambda$ is the decay constant: $\lambda$ = $\frac{\ln{2}}{T_{1/2}}$
($T_{1/2}$ is the decay  half-life).
The solution of Eq.~\ref{eq:4.2} is
\begin{equation}
\label{eq:4.3}
N = \frac{R}{\lambda} (1 - e^{-\lambda t}),
\end{equation}
and the activity $A(t)$ is related to the number of existing nuclei $N$ by
\begin{align}
\label{eq:4.4}
A(t) 	&= 
	\lambda \cdot N  \\
	&= 
	R (1 - e^{-\lambda t}) 
\end{align}
When the time $t$ is sufficiently large, then a saturation activity $A_{s}$ is reached at a given place,  
\begin{equation}
\label{eq:4.6}
A_{s} = f \cdot R_{s}
\end{equation}
The ANAIS experiment~\cite{cosmogenic-amare15} realized that the cosmic-ray neutron flux $\Phi$ can be scaled
from its sea level reference flux, as reported in Ref.~\cite{walter-thesis}, to the NaI(Tl) crystal production point
in Grand Junction, Colorado (altitude=1400 m), by a factor of $f=3.6$.  Using the measured $A_0$ activity levels
reported here and the exposure times listed in Table~\ref{crystal}, we compute a production rate R$_{s}$ at
sea level from the relation, 
\begin{equation}
\label{eq:4.7}
A_{0} = R_{s} [1 + (f (1 - e^{-\lambda t_{1}}) - 1) \cdot e^{-\lambda t_{2}}]
\end{equation} 
where the exposure time $t_{exp}$, listed in Table~\ref{crystal}, is $t_{exp}~=~t_{1}+t_{2}$.
The crystals are exposed at Alpha Spectra for a time  $t_{1}$ and exposed at sea level for a time $t_{2}$. 
We considered $t_{2}$ = 30 days as transportation duration to Y2L.
Table~\ref{production_rate} shows the production rate of cosmogenic isotopes in each NaI(Tl) crystal used
for the COSINE-100 experiment compared with measurements from ANAIS~\cite{cosmogenic-amare15,cosmogenic-villar18,anais112:2019}
and DM-Ice17~\cite{walter-thesis}, and calculations using ACTIVIA and  MENDL-2; we used v1.3 of ACTIVIA that follows the parameterization of Gordon~\cite{gordon:2004}, valid from 1~MeV to 10~GeV, for the neutron flux spectrum and MENDL-2 that contains neutron reaction data up to 100~MeV. 
The results obtained with the six NaI(Tl) crystals are in reasonable agreement with each other, except for Crystal-3 that has a complicated exposure history; 
we considered 1 year exposure time for Crystal-3 for calculation of the production rates listed in Table~\ref{production_rate}.
The production rates of $^{22}$Na in the six crystals are compatible with each other although they are larger than other measurement and calulation.  
For Crystal-4, since we do not know clearly the month and day of the powder production and the crystal delivery to Y2L it is possible to have about 60 days uncertainty for the exposure and the cooling times respectivley.

\section{Discussion on Tritium $^{3}$H and Iodine $^{129}$I}
\label{sec:5}
\begin{table*}[ht]
\begin{center}
    \caption{
    Initial activity (A$_{0}$) of $^{3}$H in the NaI(Tl) crystals derived from background fitting and the derived estimated exposure times.
    }
    \label{exposuretime_H3}
    \begin{tabular}{c|cccccc}
      \hline           
      &  Crystal-1  &  Crystal-2   &  Crystal-3  &  Crystal-4  &  Crystal-6   &  Crystal-7  \\ \hline 
      A$_{0}$ [mBq/kg] & 0.38$\pm$0.04 & 0.20$\pm$0.04  & 0.25$\pm$0.04  & 0.26$\pm$0.04  & 0.11$\pm$0.04  & 0.09$\pm$0.04  \\
      Exposure time [year] & 2.19 & 1.11 & 1.37 & 1.44 & 0.66 & 0.52 \\ \hline   
    \end{tabular}
\end{center}
\end{table*}

It is generally difficult to measure activity levels of long-lived cosmogenic isotopes,
directly from the data due to their long half-lives. This is especially the case for $^{3}$H,
which has no distinguishing $\gamma$/X-ray peak that can be exploited. 
Therefore, we simulated background spectra from $^{3}$H in the six NaI(Tl) crystals and used
the extracted spectral shapes in the data fitting, while floating their unknown fractions~\cite{cosinebg}.
We determine the initial activity A$_{0}$ of $^{3}$H by using the average activity during the first 60 days of data, obtained from the global background fitting model~\cite{cosinebg} described above,
with results shown in Table~\ref{exposuretime_H3}. From these we computed exposures time $t_{exp}$ from Eq.~(\ref{eq:4.7}). 
We assumed that the production rate of $^{3}$H at sea level is R$_{s}$~=~(83$\pm$27)~kg$^{-1}$d$^{-1}$, which was reported by ANAIS~\cite{cosmogenic-amare18}.
The resultant exposure times, listed in Table~\ref{exposuretime_H3}, are in good agreement with the time
period during which they were being produced at Alpha Spectra and undergoing delivery to Y2L, as shown in
Table~\ref{crystal}.

The presence of cosmogenic $^{129}$I was introduced  by DAMA/LIBRA with the estimated concentration of
$^{129}$I$\slash$ $^{nat}$I~=~(1.7$\pm$0.1)$\times$10$^{-13}$~\cite{Bernabei:2008yh}. 
It is used as a floating parameter in the global background fitting modeling for the COSINE-100 NaI(Tl) crystals, with resulting
values of 1.01, 1.08, 0.75, 0.72, 0.91, and 0.94~mBq/kg for Crystal-1, 2, 3, 4, 6, and 7, respectively.
These values agree well with the ANAIS result: 0.96$\pm$0.06~mBq/kg~\cite{anais0:2012}.

\section{Conclusion} 
\label{sec:conc}
We have studied background contributions from cosmogenic isotopes activated by cosmic rays in the
COSINE-100 detectors. To understand their time-dependent energy spectra we simulated responses
to decays of the most abundantly produced cosmogenic isotopes in NaI(Tl) crystals
and identified the energy regions where they make strong contributions to the crystals' background 
spectra. Based on these simulation studies we measured decay rates of the
cosmogenic isotopes using the time-dependent decrease of peaks from characteristic decays of these
isotopes. We also exploited the correlations of characteristic emissions  
in terms of time differences between sequential decays of $^{109}$Cd and double- and triple-coincidences for
$^{22}$Na-decay-induced multi-gamma final states.  From these measurements, we extrapolated the
various isotopes' activity levels to the times that they were first deployed
underground at Y2L.

With these data we estimated production rates (at sea level) for the cosmogenic isotopes that are
relevant for COSINE-100 and compared them with other experimental data and ACTIVIA/MENDL-2 calculations.
As listed in Table~\ref{production_rate}, the results from different approaches are in reasonable
agreement with each other. We extracted exposure times
using initial $^{3}$H activities determined from the COSINE-100 global background fitting
model and found results that are in  
reasonable agreement with the times reported in Table~\ref{cosmogenic}(b). 
We also quantified the unknown $^{129}$I activity level  by including it as a free-floating
parameter in the the global background fit model and found consistency with an ANAIS result.
 
This study has given us a quantitative understanding of the cosmogenic isotopes in the
NaI(Tl) crystals used for the COSINE-100 experiment. It provides important constraints on
time-dependent backgrounds in our search for a time-dependent modulation that would be a
characteristic signal for dark matter interactions~\cite{anais-modulation:2019,cosine100-modulation:2019}. 


\section*{Acknowledgments}
We thank the Korea Hydro and Nuclear Power (KHNP) Company for providing underground laboratory space at Yangyang. 
This work is supported by: the Institute for Basic Science (IBS) under project code IBS-R016-A1 and 
NRF-2016R1A2B3008343, Republic of Korea; UIUC campus research board, the Alfred P. Sloan Foundation Fellowship, 
NSF Grants No. PHY-1151795, PHY-1457995, DGE-1122492, WIPAC, the Wisconsin Alumni Research Foundation, United States; 
STFC Grant ST/N000277/1 and ST/K001337/1, United Kingdom; and Grant No. 2017/02952-0 FAPESP, CAPES Finance Code 001, Brazil.



\bibliographystyle{elsarticle-num} 





\end{document}

%% file: authors_May2019.tex
\author[d]{E.~Barbosa~de~Souza}
\author[o]{B.~J.~Park\corref{cor1}}  
\ead{pbj7363@gmail.com}
\author[c]{G.~Adhikari} 
\author[c]{P.~Adhikari\fnref{1}}
\fntext[1]{Department of Physics, Carleton University, Ottawa, Ontario, K1S 5B6, Canada}
\author[e]{N.~Carlin}
\author[f]{J.~J.~Choi} 
\author[f]{S.~Choi} 
\author[a]{M.~Djamal} 
\author[h]{A.~C.~Ezeribe} 
\author[b]{C.~Ha} 
\author[i]{I.~S.~Hahn}   
\author[b]{E.~J.~Jeon\corref{cor1}}  
\ead{ejjeon@ibs.re.kr}
\author[d]{J.~H.~Jo}  
\author[b]{W.~G.~Kang}  
\author[k]{M.~Kauer} 
\author[l]{G.~S.~Kim} 
\author[b]{H.~Kim}  
\author[l]{H.~J.~Kim}  
\author[b]{K.~W.~Kim}  
\author[b]{N.~Y.~Kim}  
\author[f]{S.~K.~Kim}  
\author[b,c,o]{Y.~D.~Kim}  
\author[b,m,o]{Y.~H.~Kim} 
\author[b]{Y.~J.~Ko} 
\author[h]{V.~A.~Kudryavtsev}
\author[b]{E.~K.~Lee} 
\author[b,o]{H.~S.~Lee}  
\author[b]{J.~Lee}  
\author[l]{J.~Y.~Lee}  
\author[b,o]{M.~H.~Lee} 
\author[o]{S.~H.~Lee} 
\author[b]{D.~S.~Leonard}
\author[h]{W.~A.~Lynch}
\author[e]{B.~B.~Manzato}  
\author[d]{R.~H.~Maruyama}
\author[h]{R.~J.~Neal} 
\author[b]{S.~L.~Olsen}  
\author[p]{H.~K.~Park} 
\author[m]{H.~S.~Park}   
\author[b]{K.~S.~Park} 
\author[e]{R.~L.~C.~Pitta}  
\author[a,b]{H.~Prihtiadi}  
\author[b]{S.~J.~Ra}  
\author[j]{C.~Rott} 
\author[b]{K.~A.~Shin} 
\author[h]{A.~Scarff}  
\author[h]{N.~J.~C.~Spooner}  
\author[d]{W.~G.~Thompson}  
\author[n]{L.~Yang} 
\author[j]{G.~H.~Yu}
\author[]{(COSINE-100 Collaboration)}

\cortext[cor1]{Corresponding authors}

\address[d]{Department of Physics, Yale University, New Haven, CT 06520, USA}
\address[o]{IBS School, University of Science and Technology (UST), Daejeon 34113, Republic of Korea} 
\address[c]{Department of Physics, Sejong University, Seoul 05006, Korea}
\address[e]{Physics Institute, University of S\~{a}o Paulo, S\~{a}o Paulo 05508-090, Brazil}
\address[f]{Department of Physics and Astronomy, Seoul National University, Seoul 08826, Korea}    
\address[a]{Department of Physics, Bandung Institute of Technology, Bandung 40132, Indonesia}
\address[h]{Department of Physics and Astronomy, University of Sheffield, Sheffield S3 7RH, United Kingdom}    
\address[b]{Center for Underground Physics, Institute for Basic Science~(IBS), Daejeon 34047, Korea} 
\address[i]{Department of Science Education, Ewha Womans University, Seoul 03760, Korea} 
\address[k]{Department of Physics and Wisconsin IceCube Particle Astrophysics Center, University of Wisconsin-Madison, Madison, WI 53706, USA}   
\address[l]{Department of Physics, Kyungpook National University, Daegu 41566, Korea}    
\address[m]{Korea Research Institute of Standards and Science, Daejeon 34113, Korea}   
\address[p]{Department of Accelerator Science, Korea University, Sejong 30019, Republic of Korea}
\address[j]{Department of Physics, Sungkyunkwan University, Suwon 16419, Korea}   
\address[n]{Department of Physics, University of Illinois at Urbana-Champaign, Urbana, IL 61801, USA}  

%% file: analysis1_v1.9.tex
\subsection{Measurement of decay rates}
\label{sec:anal1}
\begin{figure}[ht]
\begin{center}
\includegraphics[width=0.5\textwidth]{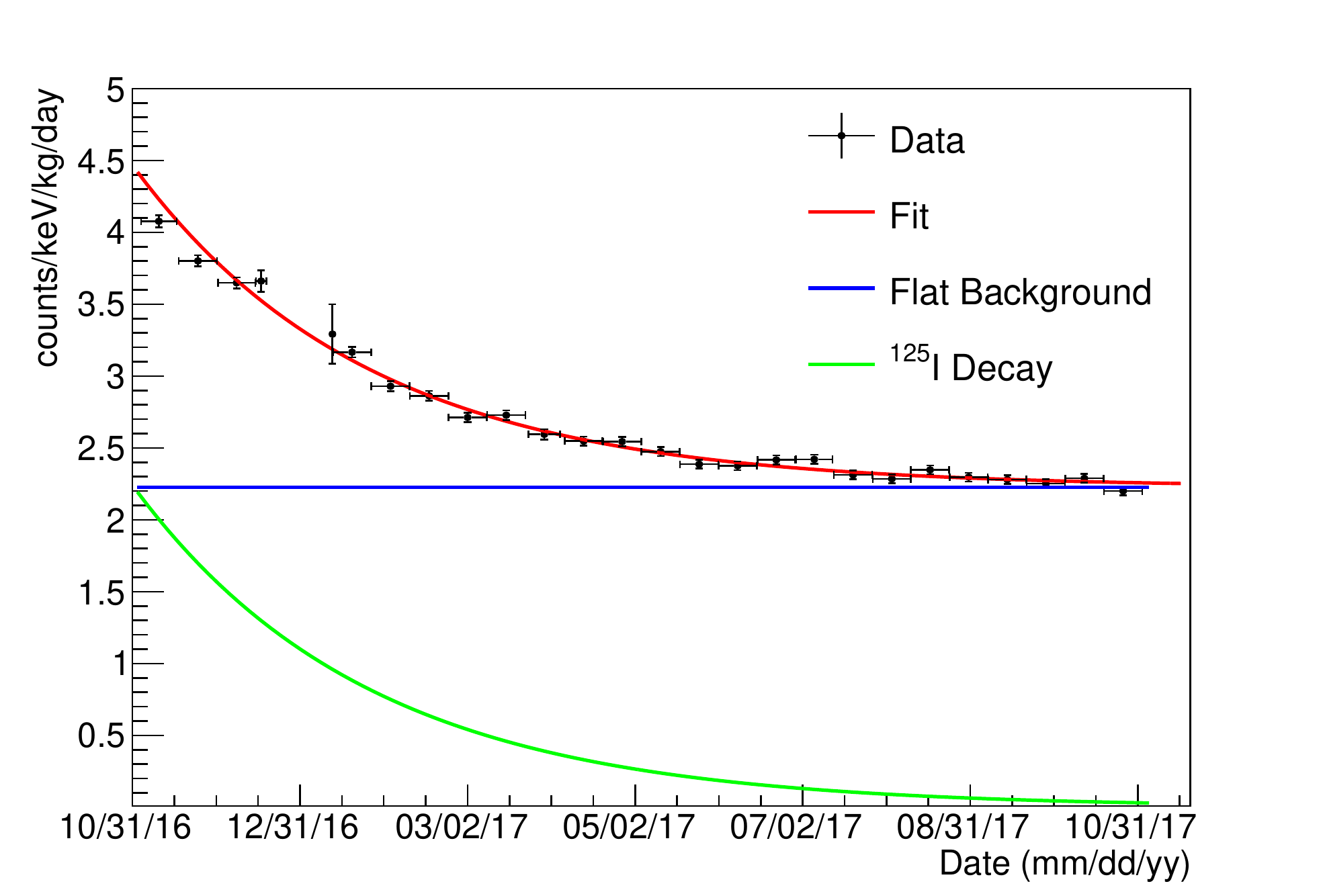}
\caption{
Average rate in 60-70 keV region of Crystal-4's single-hit spectrum over time. The rate shows a clear decrease, which can be modeled with an exponential decay, for $^{125}$I in this case, in addition to a flat background component. The data is divided in bins of 15 days. However, down times between runs can result in some bins having less statistics than others.
It is not expected to contribute to either Crystal-1 or Crystal-2 due to their relatively long cooling times underground. 
}
\label{fig:decay}
\end{center}
\end{figure}

\begin{figure}[ht]
\begin{center}
\includegraphics[width=0.5\textwidth]{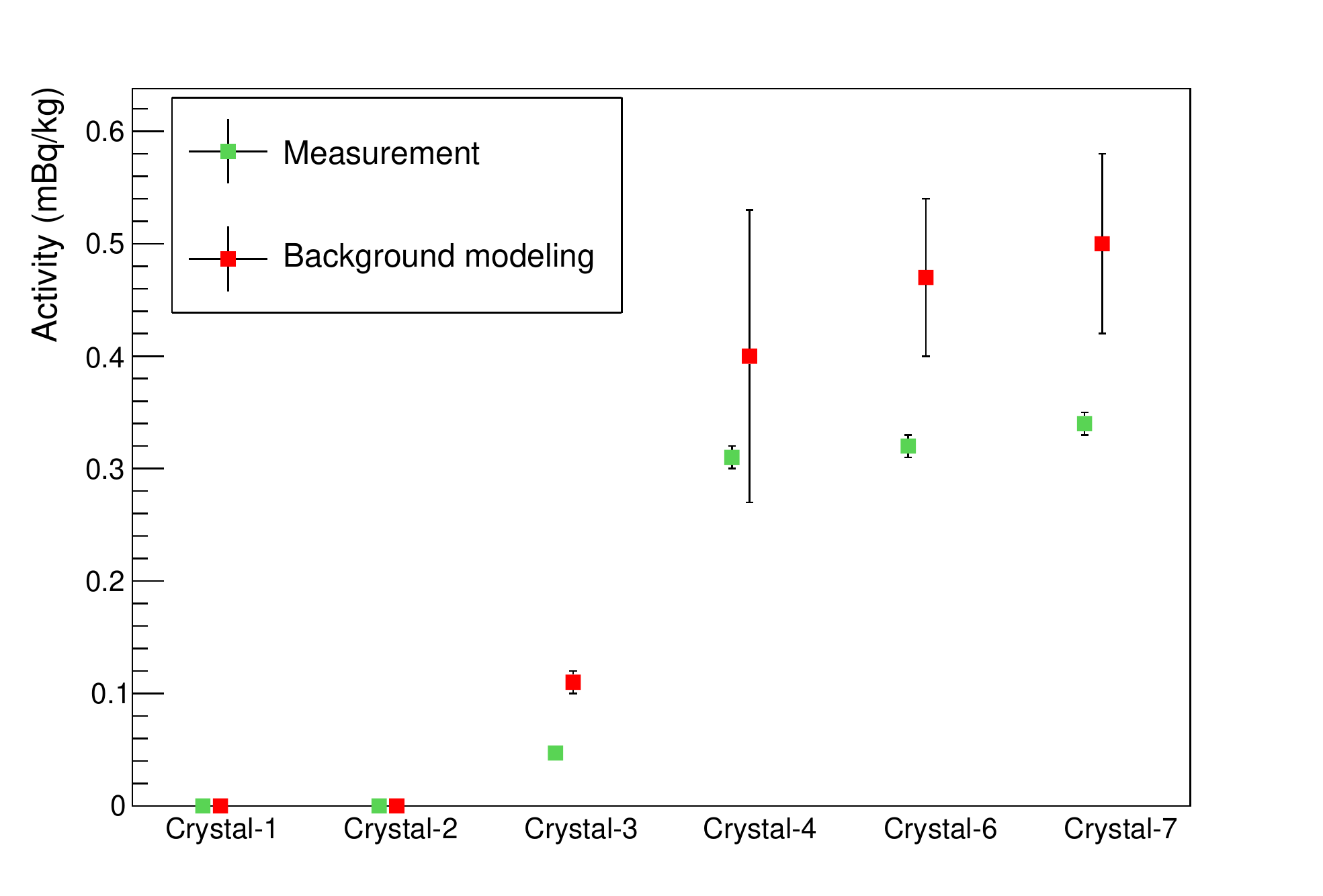}
\caption{Average activities of $^{125}$I during the first 60 days of data.  
The values measured through the decay fit method, described in the text, 
are plotted in green, while the ones measured through the background modeling~\cite{cosinebg} are shown in red.
}
\label{fig:I125}
\end{center}
\end{figure}

\begin{table*}[ht]
\begin{center}
\caption{
Initial activity A$_{0}$ (mBq/kg) of $^{125}$I in each crystal as measured by the decay rate method. This includes the statistical uncertainty.
}
\label{table:I125}
\begin{tabular}{c|cccccc}
\hline             
      &  Crystal-1  &  Crystal-2   &  Crystal-3  &  Crystal-4  &  Crystal-6   &  Crystal-7  \\ \hline 
      $^{125}$I & - & - & 9.0$\pm$0.9 & 3.4$\pm$0.1 & 5.1$\pm$0.2 & 5.4$\pm$0.2 \\
\hline            
\end{tabular}
\end{center}
\end{table*}

One way to measure the activities of the cosmogenic isotopes is through their decay rates. This measurement requires a selection of events of the specific decays studied, which can be identified by investigating the main contributions of the decay to our data spectra. Therefore, we first simulate each of the cosmogenically activated isotopes with the COSINE-100 GEANT4 package, studying their generated spectra and selecting the energy regions where they can have a significant contribution in comparison to the flat background. 

\begin{figure*}[!ptb]
\begin{center}
\begin{tabular}{ccc}
\includegraphics[width=0.3\textwidth, trim = {1.3cm 0 1.5cm 0}]{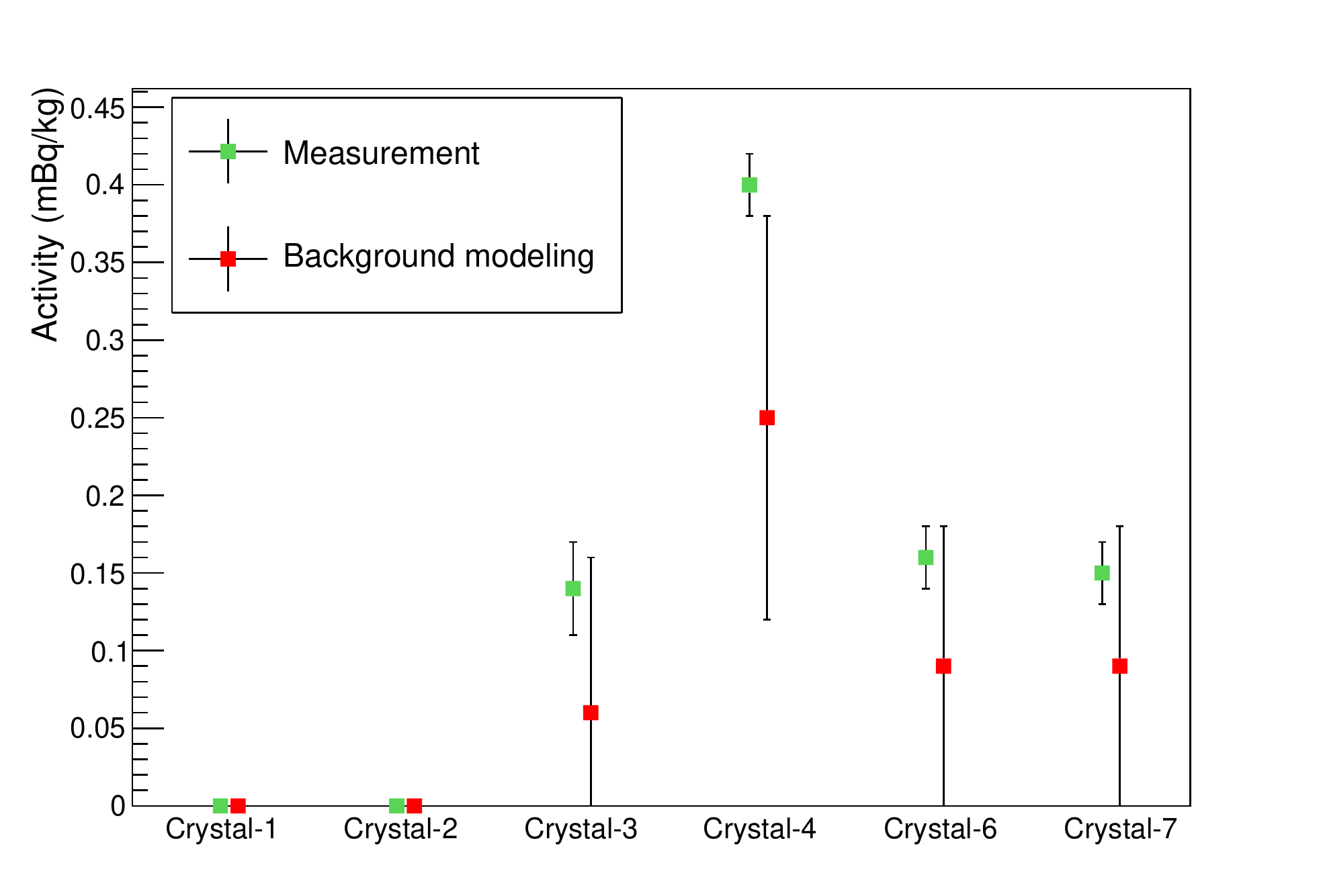} &
\includegraphics[width=0.3\textwidth, trim = {1.3cm 0 1.5cm 0}]{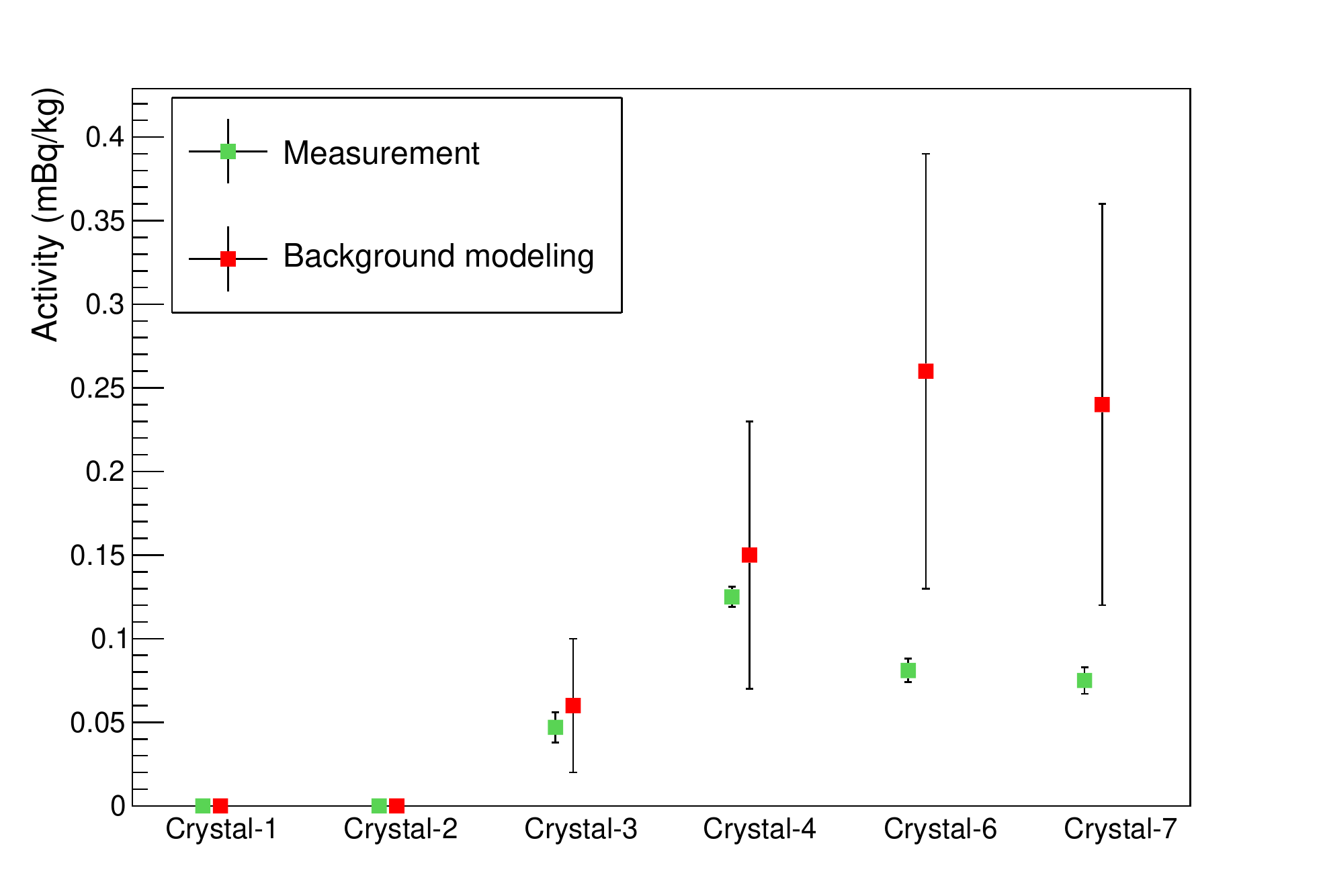} &
\includegraphics[width=0.3\textwidth, trim = {1.3cm 0 1.5cm 0}]{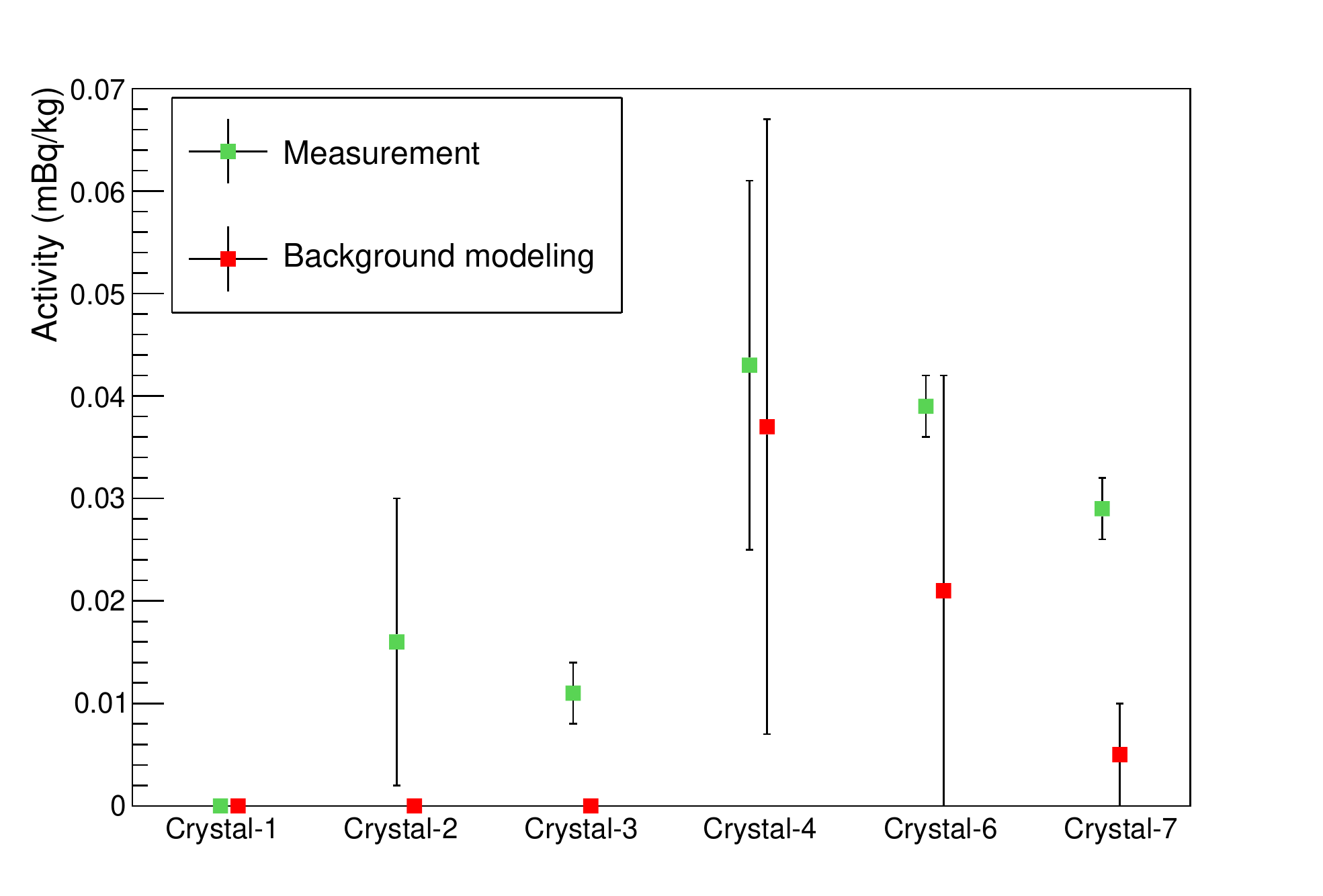} \\
(a) & (b) & (c) \\
\end{tabular}
\caption{
Average activities of (a) $^{121m}$Te, (b) $^{127m}$Te, and (c) $^{113}$Sn during the first 60 days of data in six crystals. The values measured through the decay fit method, described in the text, 
are plotted in green, while the ones measured through the background modeling~\cite{cosinebg} are shown in red.
}
\label{fig:Te}
\end{center}
\end{figure*}

Once the regions where the main contributions from each cosmogenically activated isotope are identified, we look into the  decay rates for each of them, integrating the rates over the specific energy ranges. We fit the decay rate over time for each component. Depending on the energy region selected for each cosmogenic, the decay rate can be fitted by a constant and one or more exponential functions, as following:
\begin{equation}
A + B \cdot e^{\frac{-ln(2)(t-t_0)}{C}}~,
\end{equation}
where $t_0$ is the initial time,
$A$ is the expected flat background rate, B is the rate in $t_0$, 
and $C$ is the half-life, a constant of the fit.  
Fig.~\ref{fig:decay} shows an example of decay rate modeling with the units given in dru~(counts/day/kg/keV). 

The amplitude of the exponential ($B$) can be used to calculate the activity (in Bq/kg) of the cosmogenic isotope at the indicated initial time:
\begin{equation}
A_0 = \frac{\Delta E \times B}{86400\times f_{\Delta E}}~,
\end{equation}
where $f_{\Delta E}$ is the fraction of the events from that cosmogenic depositing energy in the specified integration region, which can be calculated from the simulated spectra. 
$\Delta E$s for each isotope are 60--70~keV in single hit for $^{125}$I, 20--40~keV in multiple hit for $^{121m}$Te, 
80--94~keV in single hit for $^{127m}$Te, and 20--30~keV in single hit for $^{113}$Sn.

\subsubsection{Iodine $^{125}$I}
Since iodine is one of the main components in the crystal, a significant amount of $^{125}$I is activated. However, the half-life of this isotope is short with $T_{1/2}$~=~59.4 days.
We define the integration region for this isotope as $60$ and $70$ keV of the single-hit spectrum, as shown in Fig.~\ref{fig:decay}; it decays to an isomeric state of $^{125m}$Te by electron capture, producing 31.7 keV emissions for K-shell electrons, which is followed by the emission of an 35.5 keV gamma ray from the isomer transition of $^{125m}$Te.  Although there is a contribution from $^{121m}$Te in this region, it is very small due to the low activity of this isotope and more importantly, due to the very small fraction of total $^{121m}$Te events that actually deposit energy in this region. Therefore, the contribution of $^{121m}$Te in the $60$ to $70$ keV region is insignificant
, and the amounts of activated $^{125}$I in the crystals during their deployment at Y2L can be found in Table~\ref{table:I125}. 

Fig.~\ref{fig:I125} shows the difference between the measured activities of $^{125}$I for the six crystals analyzed.  
The different amounts of $^{125}$I are related to the cooling time of the crystals before the start of data taking. We also compare the activities calculated in the background fit~\cite{cosinebg} to those measured through this method. 

\subsubsection{Tellurium $^{121m}$Te, $^{127m}$Te and Tin $^{113}$Sn }
\label{sec:anal1.2}
\begin{table*}[ht]
\begin{center}
\caption{
Initial activity A$_{0}$ (mBq/kg) of $^{121m}$Te, $^{127m}$Te, and $^{113}$Sn in each crystal as measured by the decay rate method.
This includes the statistical uncertainty.
}
\label{table:Te}
\begin{tabular}{c|cccccc}
\hline             
      &  Crystal-1  &  Crystal-2   &  Crystal-3  &  Crystal-4  &  Crystal-6   &  Crystal-7  \\ \hline
      $^{121m}$Te & - & - & 0.90$\pm$0.16 & 0.89$\pm$0.06 & 0.44$\pm$0.07 & 0.41$\pm$0.07 \\  
      $^{127m}$Te & - & - & 0.87$\pm$0.16 & 0.48$\pm$0.03 & 0.38$\pm$0.04 & 0.35$\pm$0.04 \\ 
      $^{113}$Sn & - & - & 0.16$\pm$0.05 & 0.15$\pm$0.06 & 0.16$\pm$0.01 & 0.12$\pm$0.01 \\ \hline           
\end{tabular}
\end{center}
\end{table*}

\begin{figure*}[ht]
  \begin{center}
    \begin{tabular}{cc}
      \includegraphics[width=0.5\textwidth]{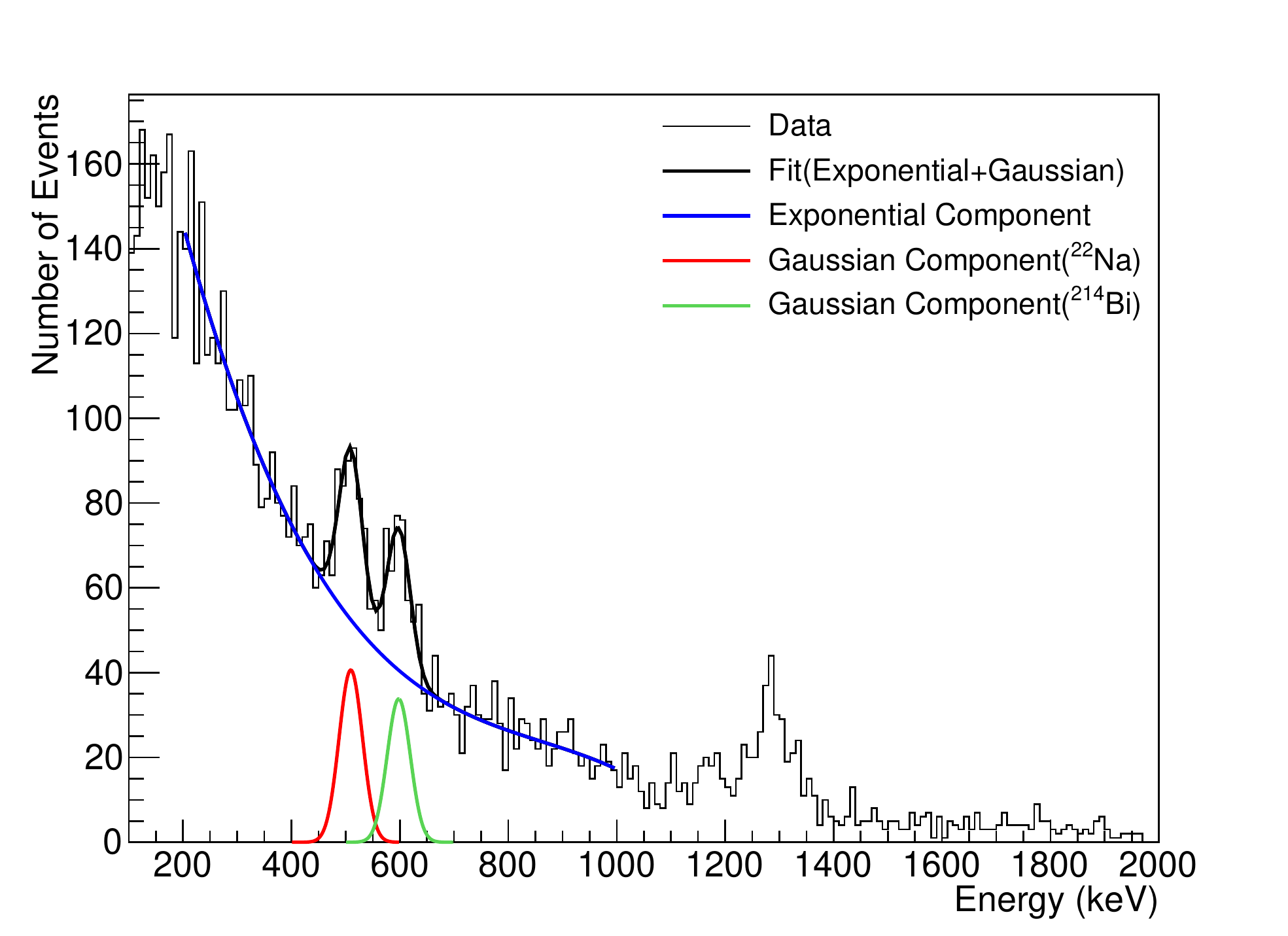} &
      \includegraphics[width=0.5\textwidth]{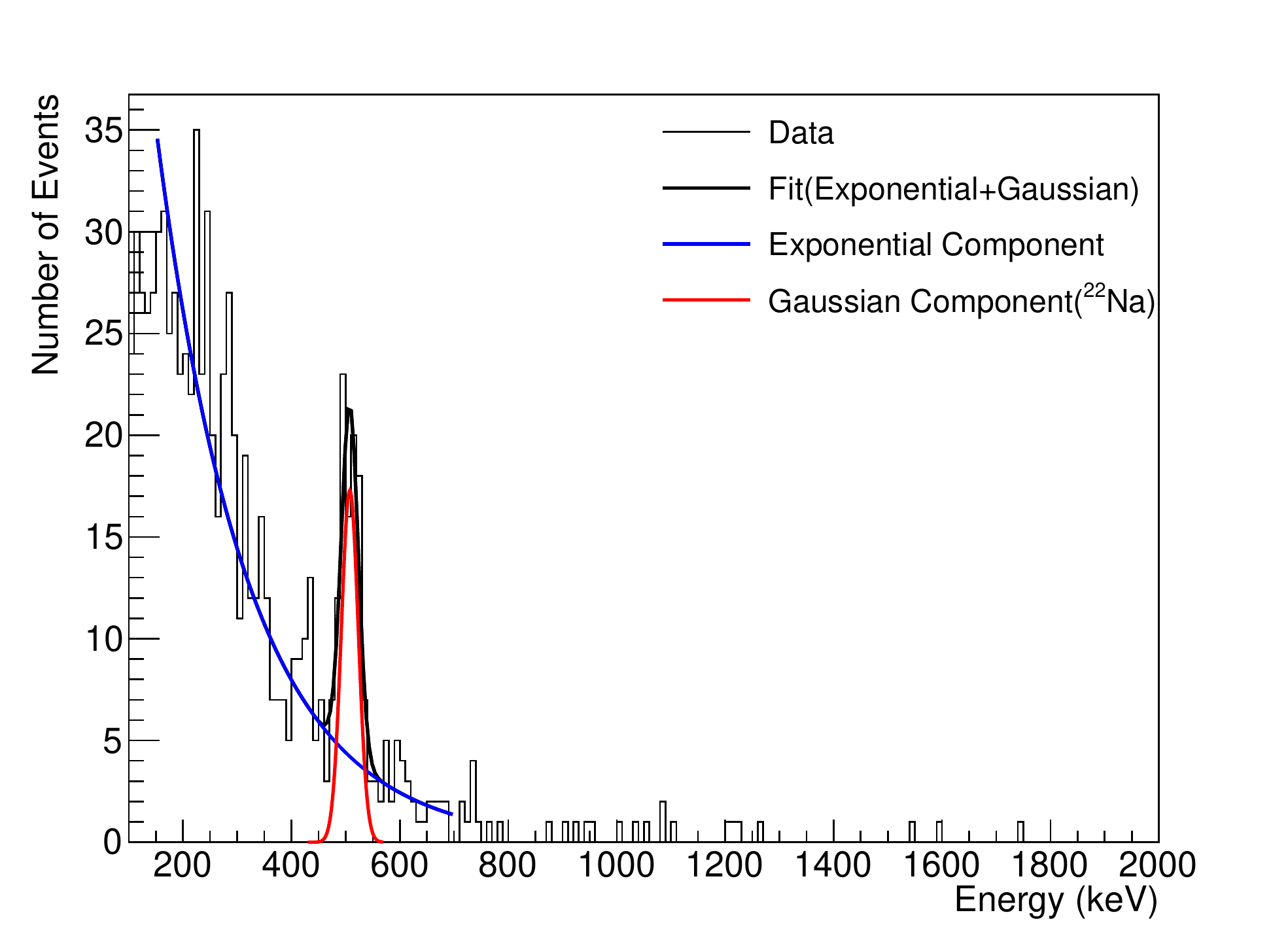} \\
      (a) & (b)\\
    \end{tabular}
  \end{center}
  \caption{(a) Energy spectrum in another crystal in coincidence with a signal in the (650–1000) keV energy interval in Crystal-3. 
  In the fitting, it represents the peak at 609~keV from $^{214}$Bi  in green and the peak at 511~keV from $^{22}$Na, activated in Crystal-3, in red.   
(b) Energy spectrum in another crystal contributed by triple coincidence events.  
The red line represents the peak at 511~keV from $^{22}$Na in Crystal-3.
The same plots for all the other crystals are similar to each other.
}
  \label{ref:fig7}
\end{figure*}

\begin{itemize}
  \item[$\bullet$] 
The line chosen to investigate $^{121m}$Te is the one between $20$ to $40$ keV in the multiple-hit spectrum, contributed by 30~keV emissions from the  $^{121m}$Te decay via electron capture. 
This line is dominant in that region, unlike the $^{121m}$Te lines in the single-hit spectra. The method used is the same as described above.
  \item[$\bullet$]
As we can see from the simulated spectra, $^{127m}$Te has only one peak at 88.3~keV in the single-hit spectrum, 
contributed by the de-excitation of $^{127m}$Te.
However, there are other components that can have significant contributions in that energy region as well, such as $^{109}$Cd and $^{121m}$Te: 
a 88~keV emission from the isomer transition of $^{109m}$Ag and another one at 81.8~keV from the de-excitation of $^{121m}$Te.
The contribution coming from $^{121m}$Te can be calculated based on the measurement from the multiple-hit spectrum, as described above. 
The contribution from $^{109}$Cd, however, is calculated based on another study, which will be described in section~\ref{sec:anal2.2}. 
Both of these are added to the fitting function, allowing the measurement of $^{127m}$Te, activity. 
  \item[$\bullet$]
The electron capture decay of $^{113}$Sn produces 28~keV emissions.
The method used is the same as described above.
The peaks at 25.5~keV and 30~keV contributed by $^{109}$Cd and $^{121m}$Te can be calculated, as described above, and added to the fitting function, allowing the measurement of $^{113}$Sn.
\end{itemize}
  
Fig.~\ref{fig:Te}(a), (b) show the differences between the measured activities of $^{121m}$Te and $^{127m}$Te for the six crystals analyzed. We also compare the activities calculated in the background fit to those measured through this method.
As listed in Tables~\ref{table:I125} and \ref{table:Te}, 
the initial activities when crystals were deployed at Y2L were derived for $^{125}$I, $^{121m}$Te, $^{127m}$Te, and $^{113}$Sn in each crystal. 

%% file: analysis2_v1.9.tex
\subsection{Long-lived isotopes}
\label{sec:anal2}
\subsubsection{Sodium $^{22}$Na}
\begin{figure}[!b]
  \begin{center}
    \begin{tabular}{c}
      \includegraphics[width=0.5\textwidth]{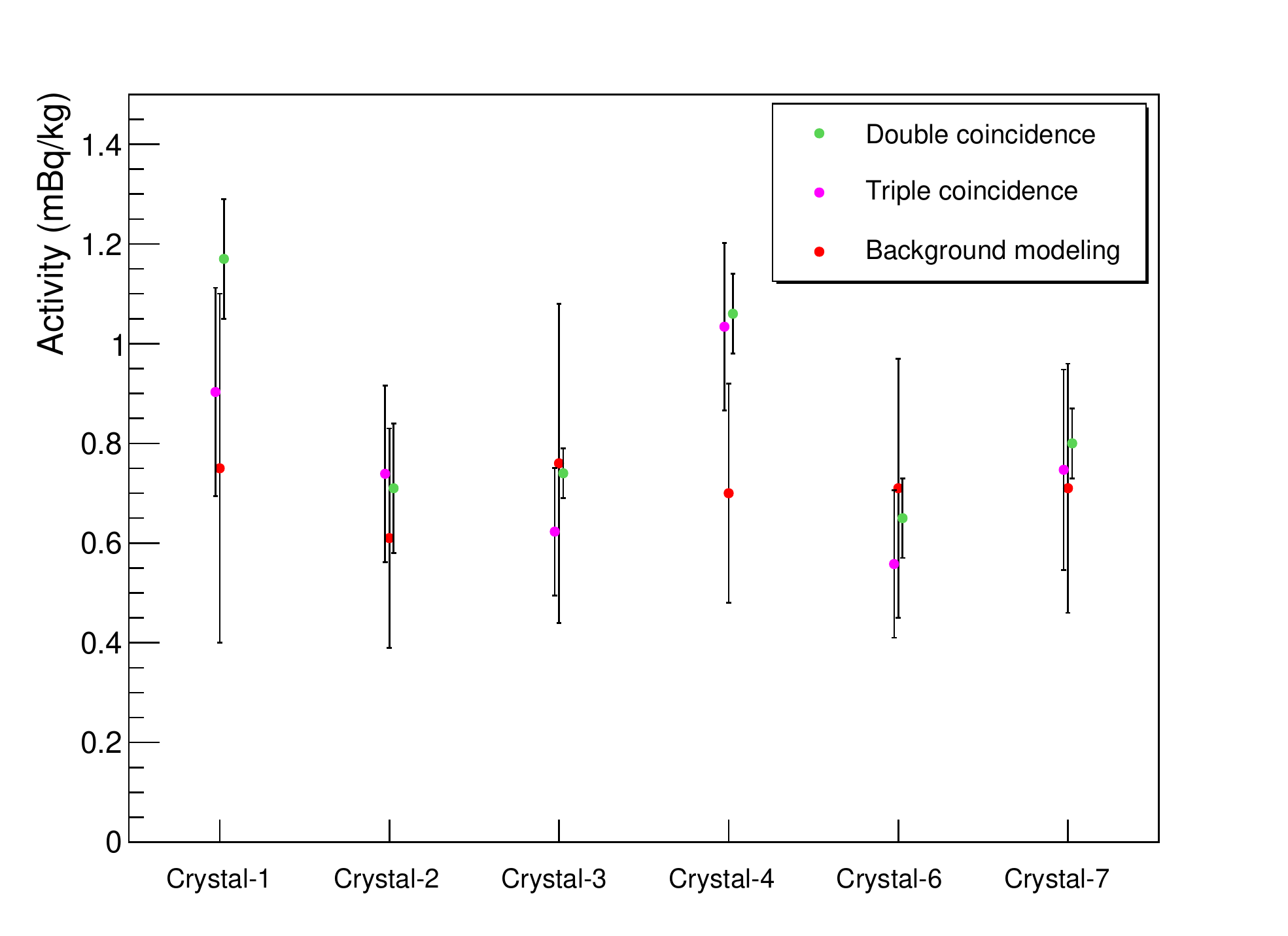}
    \end{tabular}
  \end{center}
  \caption{Average activities of $^{22}$Na during the first 60 days of data in six crystals.
  The values measured through the double/triple coincidence methods, described in the text, 
  are plotted in green/pink, while the ones measured through the background modeling~\cite{cosinebg} are shown in red.}
  \label{fig:figure8}
\end{figure}

\begin{table*}[ht]
\begin{center}
    \caption{
    Initial activity A$_{0}$ (mBq/kg)  of $^{22}$Na as measured by double and triple coincidences in each crystal.
    This includes the statistical uncertainty.
    }
    \label{tab:table4}
    \begin{tabular}{c|cccccc}
      \hline            
     &  Crystal-1  &  Crystal-2   &  Crystal-3  &  Crystal-4  &  Crystal-6   &  Crystal-7  \\ \hline
     double coincidence & 2.59$\pm$0.27 & 1.46$\pm$0.27 & 0.99$\pm$0.07 & 1.20$\pm$0.09 & 0.73$\pm$0.09 & 0.93$\pm$0.08 \\
     triple coincidence & 2.0$\pm$0.4 & 1.52$\pm$0.37 & 0.84$\pm$0.18 & 1.17$\pm$0.19 & 0.65$\pm$0.14 & 0.87$\pm$0.23 \\ \hline              
    \end{tabular}
\end{center}
\end{table*}
 
The decays of $^{22}$Na (Q-value~=~2.84~MeV) to $^{22}$Ne$^{*}$ proceed via $\beta^+$ emission (90.3\%) or electron capture (9.6\%) 
with 3.75~yr mean lifetime, followed by $^{22}$Ne$^{*}$ transitioning to the stable $^{22}$Ne isotope via
the emission of a 1274.6~keV $\gamma$-ray with a 5.3~ps mean lifetime. 
The electron capture of $^{22}$Na from K-shell produces 0.87~keV emissions. 
As a result, $\sim$10\% of the $^{22}$Na decay will simultaneously produce a 1274.6~keV $\gamma$-ray and 0.9~keV emissions. 
In the case of $\beta^+$ decay, the final-state positron immediately annihilates to two 511~keV $\gamma$-rays.
If one of the two 511 $\gamma$-rays escapes the crystal, the remaining energy deposited in the crystal
will be substantially greater than 650~keV. 
Figure~\ref{ref:fig7}(a) shows the energy spectrum in another crystal in coincidence with a signal in the (650--1000)~keV energy interval in Crystal-3, which are called {\it double coincidence} events.
The $^{22}$Na $\beta^{+}$ decay events show up  
as the peak at 511~keV (red color).

Since the eight NaI(Tl) crystal assemblies are immersed in the scintillating liquid (LS), as described in
section~\ref{sec:2}, we can also identify $^{22}$Na decay events in which the 1274.6~keV gamma-ray
converts in the LS in coincidence with two 511~keV signals in two crystals. 
These are referred to as {\it triple coincidence} events.
Figure 7(b) shows the peak at 511 keV (red color) in another crystal, contributed by triple coincidence events, while a signal is in the (650–1000) keV energy interval in Crystal-3.

We used the time-dependent reduction of the peak at 511 keV, contributed by the double/triple coincidences of the 1.7 year data divided in bins of 60 days, 
to extrapolate the activity at the indicated initial time with the relation: 
\begin{equation}
  A_0 = \frac{N}{m \cdot t \cdot \epsilon}~,
\end{equation}
where $N$ is the number of events, $\epsilon$ is the detection efficiency obtained from a Monte Carlo simulation,
$m$ is the mass of the NaI(Tl) crystal, and $t$ is the time of the measurement. 

Figure~\ref{fig:figure8} shows the measured activities of $^{22}$Na for the six crystals analyzed
by these methods, compared with the activities determined from the global background fit. The
initial activities, when the crystals were first deployed at Y2L, are listed in Table~\ref{tab:table4}.

\subsubsection{Cadmium $^{109}$Cd}
\label{sec:anal2.2}
\begin{figure}[ht]
  \begin{center}
    \begin{tabular}{c}
      \includegraphics[width=0.5\textwidth]{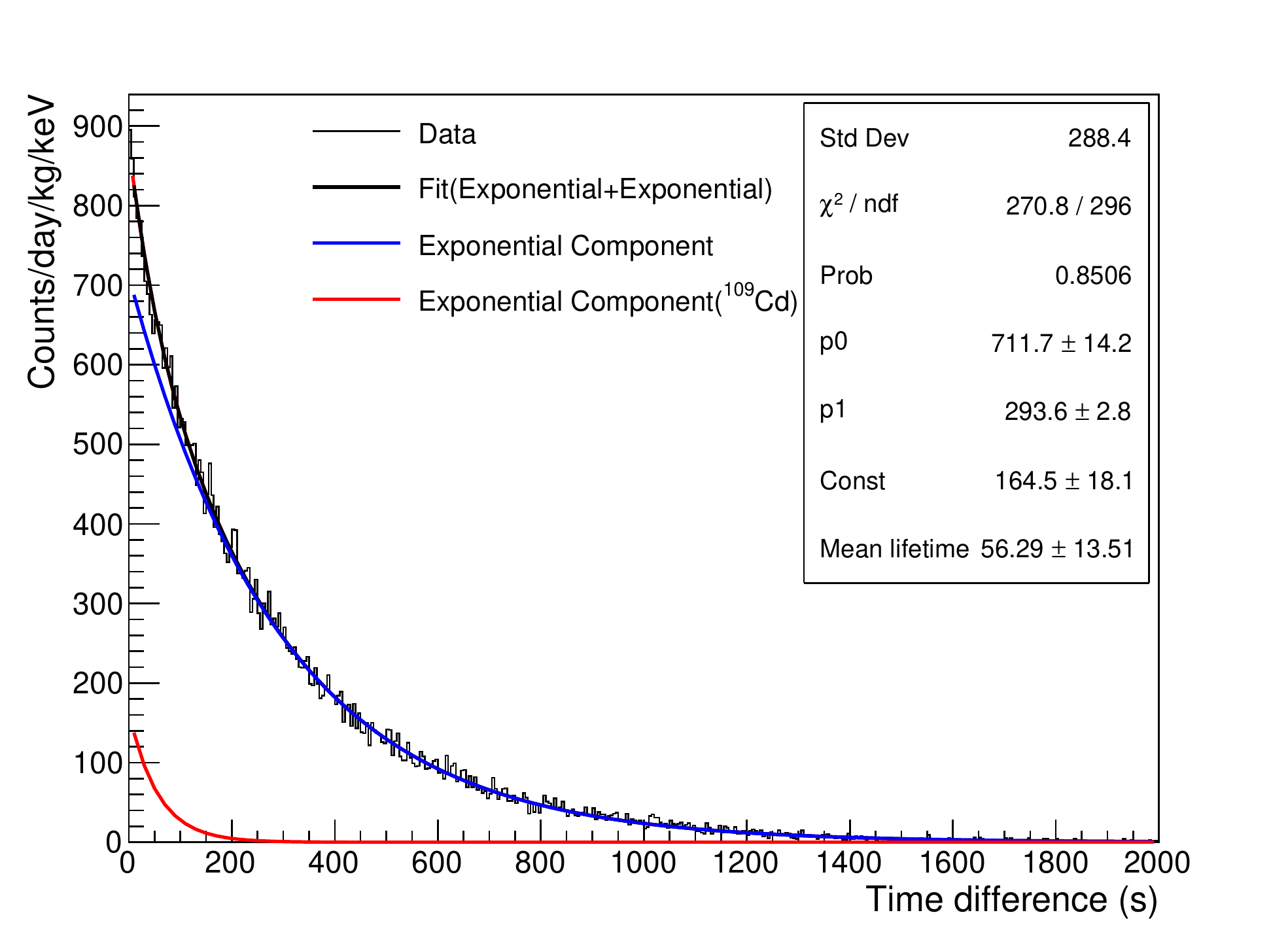} 
    \end{tabular}
  \end{center}
  \caption{Distribution of the time difference between 25.5~keV and an 88~keV signals, selected within three Gaussian width of the peaks in Crystal-3
    , fitted with the sum of two exponential decay functions. The $^{109}$Cd contribution with lifetime 56~s is shown in red.
    The parameter Const is an amplitude of the exponential decay of $^{109}$Cd. 
    The parameters p0 and p1 are an amplitude and a mean lifetime of the exponential decay of the background contribution.
    The same plots for all the other crystals are similar to each other. 
  }
  \label{fig:figure9}
\end{figure}

\begin{figure}[ht]
  \begin{center}
    \begin{tabular}{c}
      \includegraphics[width=0.5\textwidth]{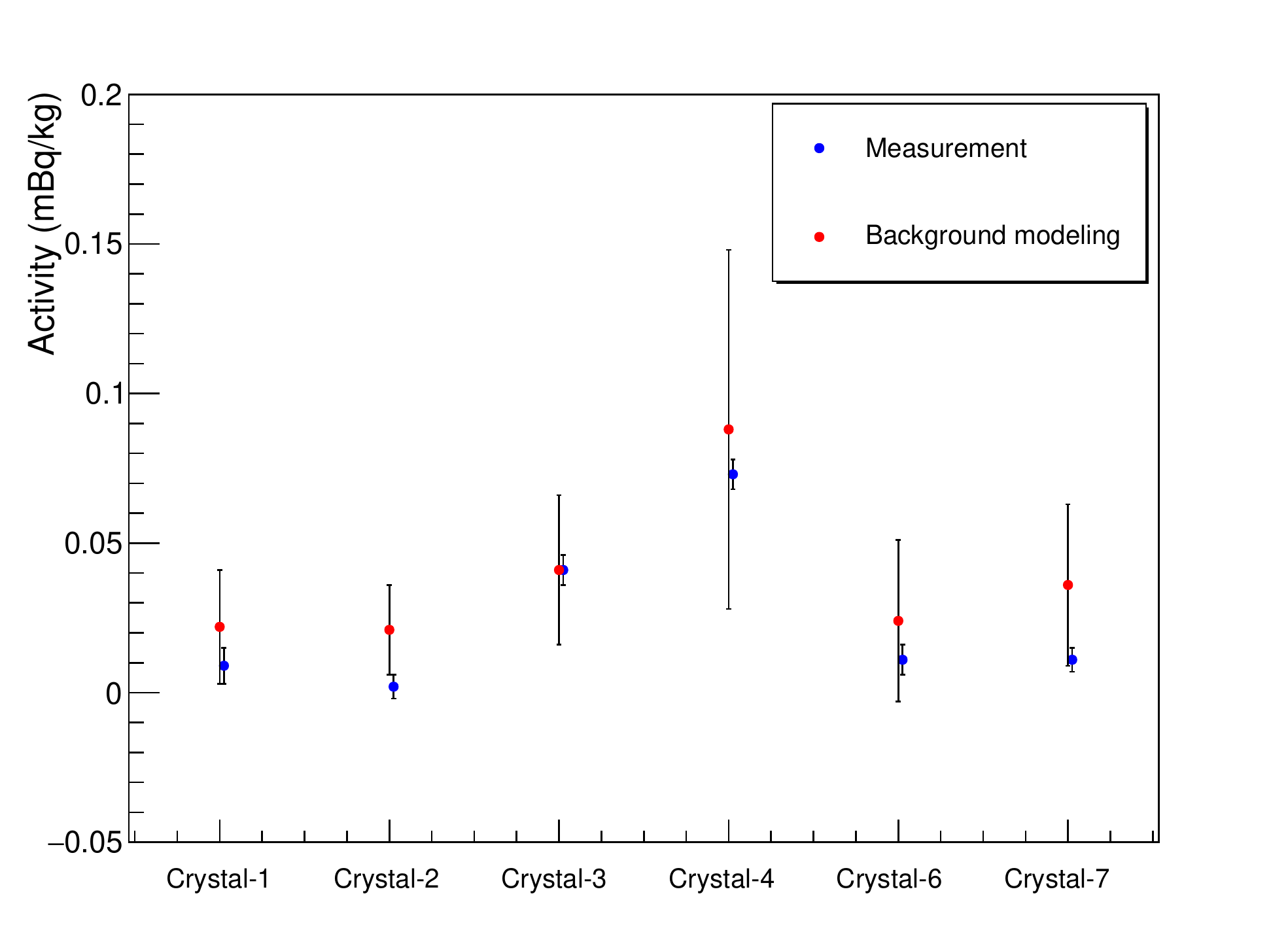}
    \end{tabular}
  \end{center}
  \caption{Average activities of $^{109}$Cd during the first 60 days of data in six crystals.
   The values measured through this method, described in the text, 
   are plotted in blue, while the ones measured through the background modeling~\cite{cosinebg} are shown in red.}
  \label{fig:figure10}
\end{figure}

\begin{table*}[ht]
\begin{center}
    \caption{Initial activity A$_{0}$ (mBq/kg) of $^{109}$Cd in each crystal. This includes the statistical uncertainty.
    }
    \label{tab:table5}
    \begin{tabular}{c|cccccc}
      \hline             
      &  Crystal-1  &  Crystal-2   &  Crystal-3  &  Crystal-4  &  Crystal-6   &  Crystal-7  \\ \hline
     $^{109}$Cd ($\times$10$^{-2}$) & 4.6$\pm$3.0 & 0.9$\pm$1.8 & 7.5$\pm$0.9 & 9.2$\pm$0.7 & 1.5$\pm$0.7 & 1.5$\pm$0.5 \\ \hline        
    \end{tabular}
\end{center}
\end{table*}

The cosmogenic isotope $^{109}$Cd decays via electron capture to an isomeric state of $^{109m}$Ag, with a prompt energy deposit of 25.5~keV, the binding energy of the Ag K-shell electron. 
This is followed by an emission of 88~keV from the isomer transition of $^{109m}$Ag that has a mean lifetime of 57.4~s. 
From the time interval distribution between 25.5~keV and 88~keV signals, selected within three Gaussian width of the peaks in the same crystal, 
we extract the level of $^{109}$Cd from a fit with two exponential decay functions.
As discussed in section~\ref{sec:anal1.2} there are significant contributions from $^{127m}$Te and $^{121m}$Te around 88~keV and $^{113}$Sn and $^{121m}$Te around 25~keV, which dominate the blue curve in Figure~\ref{fig:figure9}.
The fitted mean lifetime, 56$\pm$14~s, from the exponential curve in red, as can be seen in Figure~\ref{fig:figure9}, is consistent with the mean lifetime of
57.4~s from the isomer transition of $^{109m}$Ag.

  We determined the $^{109}$Cd activity rates in mBq/kg from these measurements: Fig.~\ref{fig:figure10} shows the
  measured activity levels for the six crystals analyzed through this method, compared with the activities determined
  from the global background fit.  The crystal activity levels when they were first deployed at Y2L
  are listed in Table~\ref{tab:table5}.

%% file: cosine-cosmogenics.bbl
\begin{thebibliography}{10}
\expandafter\ifx\csname url\endcsname\relax
  \def\url#1{\texttt{#1}}\fi
\expandafter\ifx\csname urlprefix\endcsname\relax\def\urlprefix{URL }\fi
\expandafter\ifx\csname href\endcsname\relax
  \def\href#1#2{#2} \def\path#1{#1}\fi

\bibitem{jungman96}
G.~Jungman, A.~Kamionkowski, G.~Griest, Phys. Rep. 267 (1996) 195.

\bibitem{gaitskell04}
R.~Gaitskell, Annu. Rev. Nucl. Part. Sci. 54 (2004) 315.

\bibitem{Bernabei:2013xsa}
R.~Bernabei, et~al., {Final model independent result of DAMA/LIBRA-phase1},
  Eur. Phys. J. C 73 (2013) 2648.
\newblock \href {http://dx.doi.org/10.1140/epjc/s10052-013-2648-7}
  {\path{doi:10.1140/epjc/s10052-013-2648-7}}.

\bibitem{Bernabei:2018yyw}
R. Bernabei, et~al., {First model independent results from DAMA/LIBRA--Phase2}, Nucl. Phys. At. Energy 19 (2018) 307,
arXiv:1805.10486 [astro-ph.IM].

\bibitem{sckim12}
S.~C. Kim, et~al., {New limits on interactions between weakly interacting
  massive particles and nucleons obtained with CsI(Tl) crystal detectors},
  Phys. Rev. Lett. 108 (2012) 181301.
\newblock \href {http://dx.doi.org/10.1103/PhysRevLett.108.181301}
  {\path{doi:10.1103/PhysRevLett.108.181301}}.

\bibitem{PhysRevLett.118.021303}
D.~S. Akerib, et~al.,
  \href{https://link.aps.org/doi/10.1103/PhysRevLett.118.021303}{{Results from
  a search for dark matter in the complete LUX exposure}}, Phys. Rev. Lett. 118
  (2017) 021303.
\newblock \href {http://dx.doi.org/10.1103/PhysRevLett.118.021303}
  {\path{doi:10.1103/PhysRevLett.118.021303}}.
\newline\urlprefix\url{https://link.aps.org/doi/10.1103/PhysRevLett.118.021303}

\bibitem{Cui:2017}
Xiangyi~Cui, et~al., {(PandaX-II Collaboration) Dark matter results from 54-ton-day exposure of PandaX-II experiment}, 
Phys. Rev. Lett. 119, 181302 (2017).  

\bibitem{Aprile:2018}
E.~Aprile, et~al., {Dark matter search results from a one ton-year exposure of XENON1T},
Phys. Rev. Lett. 121~(11) (2018) 111302.

\bibitem{DarkSide50:2018}
P.~Agnes et~al., {(DarkSide Collaboration) Low-mass dark matter search with the DarkSide-50 experiment}, Phys. Rev. Lett. 121, 081307 (2018)

\bibitem{SuperCDMS:2018}
R.~Agnese, et~al., {(SuperCDMS Collaboration) Results from the super cryogenic dark matter search experiment at Soudan}, 
Phys. Rev. Lett. 120, 061802 (2018)

\bibitem{Choi:2015ara}
K.~Choi, et~al., {Search for neutrinos from annihilation of captured low-mass
  dark matter particles in the Sun by Super-Kamiokande}, Phys. Rev. Lett.
  114~(14) (2015) 141301.
\newblock \href {http://dx.doi.org/10.1103/PhysRevLett.114.141301}
  {\path{doi:10.1103/PhysRevLett.114.141301}}.

\bibitem{Baum:2018ekm}
S.~Baum, K.~Freese, C.~Kelso, {Dark matter implications of DAMA/LIBRA-phase2
  results}, Phys. Lett. B789 (2019) 262-269. 

\bibitem{PhysRevD.33.3495}
A.~K. Drukier, K.~Freese, D.~N. Spergel,
  \href{https://link.aps.org/doi/10.1103/PhysRevD.33.3495}{Detecting cold
  dark-matter candidates}, Phys. Rev. D 33 (1986) 3495--3508.
\newblock \href {http://dx.doi.org/10.1103/PhysRevD.33.3495}
  {\path{doi:10.1103/PhysRevD.33.3495}}.
\newline\urlprefix\url{https://link.aps.org/doi/10.1103/PhysRevD.33.3495}

\bibitem{Savage:2006qr}
C.~Savage, K.~Freese, P.~Gondolo, {Annual modulation of dark matter in the
  presence of streams}, Phys. Rev. D74 (2006) 043531.
\newblock \href {http://dx.doi.org/10.1103/PhysRevD.74.043531}
  {\path{doi:10.1103/PhysRevD.74.043531}}.

\bibitem{Freese:2012xd}
K.~Freese, M.~Lisanti, C.~Savage, {Colloquium: Annual modulation of dark
  matter}, Rev. Mod. Phys. 85 (2013) 1561.
\newblock \href {http://dx.doi.org/10.1103/RevModPhys.85.1561}
  {\path{doi:10.1103/RevModPhys.85.1561}}.

\bibitem{deSouza:2016fxg}
E.~Barbosa~de Souza, et~al., {First search for a dark matter annual modulation
  signal with NaI(Tl) in the Southern Hemisphere by DM-Ice17}, Phys. Rev. D
  95~(3) (2017) 032006.
\newblock \href {http://dx.doi.org/10.1103/PhysRevD.95.032006}
  {\path{doi:10.1103/PhysRevD.95.032006}}.

\bibitem{Coarasa:2019}
I.~Coarasa et~al., {ANAIS-112 sensitivity in the search for dark matter annual modulation},
Eur. Phys. J. C. 79 (2019) 233.

\bibitem{sabre:2019}
M.~Antonello et~al., {The SABRE project and the SABRE Proof-of-Principle},
Eur. Phys. J. C 79 (2019) 363.

\bibitem{Fushimi:2015sew}
K.~Fushimi, et~al., {Dark matter search project PICO-LON}, J. Phys. Conf. Ser.
  718~(4) (2016) 042022.
\newblock \href {http://dx.doi.org/10.1088/1742-6596/718/4/042022}
  {\path{doi:10.1088/1742-6596/718/4/042022}}.

\bibitem{Angloher:2016}
G.~Angloher et.~al., {The COSINUS project: perspectives of a NaI scintillating calorimeter for dark matter search},
Eur. Phys. J. C 76 (2016) 441

\bibitem{Adhikari:2017esn}
G.~Adhikari, et~al., {Initial performance of the COSINE-100 Experiment}, 
Eur. Phys. J. C 78 (2018) 107.
\newblock \href {http://dx.doi.org/10.1140/epjc/s10052-018-5590-x}
  {\path{doi:10.1140/epjc/s10052-018-5590-x}}.

\bibitem{Adhikari:2018ljm}
G.~Adhikari, et~al., {An experiment to search for dark-matter interactions
  using sodium iodide detectors}, Nature 564~(7734) (2018) 83--86.
\newblock \href {http://dx.doi.org/10.1038/s41586-018-0739-1}
  {\path{doi:10.1038/s41586-018-0739-1}}.

\bibitem{cosinebg}
P.~Adhikari, et~al., {Background model for the NaI(Tl) crystals in COSINE-100},
  Eur. Phys. J. C 78 (2018) 490.
\newblock \href {http://dx.doi.org/10.1140/epjc/s10052-018-5970-2}
  {\path{doi:10.1140/epjc/s10052-018-5970-2}}.

\bibitem{activia}
Y.~R. J.J.~Back, Activia: Calculation of isotopes production cross-sections and
  yields, Nucl. Instrum. Meth. A 586 (2008) 286--294.

\bibitem{mendl-2}
Y.~N.~Shubin, V.~P.~Lunev, A.~Y.~Konobeyev, A.~I.~Kitjuk, 
{Cross-section data library MENDL-2 to study activation as transmutation of materials irradiated by nucleons of intermediate energies}, 
International Atomic Energy Agency (1995).  
\newline\urlprefix\url{http://www-nds.iaea.org/publications/iaea-nds/iaea-nds-0136.htm}

\bibitem{Agostinelli:2002hh}
S.~Agostinelli, et~al., {GEANT4: A Simulation toolkit}, Nucl. Instrum. Meth. A
  506 (2003) 250.
\newblock \href {http://dx.doi.org/10.1016/S0168-9002(03)01368-8}
  {\path{doi:10.1016/S0168-9002(03)01368-8}}.

\bibitem{DDEP}
\href{http://www.nucleide.org/DDEP_WG/DDEPdata.htm}{Decay Data Evaluation Project,} 
\newline\urlprefix\url{http://www.nucleide.org/DDEP\_WG/DDEPdata.htm}

\bibitem{Ohya:2010}
S.~Ohya, Nucl. Data Sheets 111, 1619 (2010)

\bibitem{LNDS}
\href{WWW Table of Radioactive Isotopes, http://nucleardata.nuclear.lu.se/toi/}{WWW Table of Radioactive Isotopes,}
\newline\urlprefix\url{http://nucleardata.nuclear.lu.se/toi/}

\bibitem{walter-thesis}
W.~C. Pettus, Cosmogenic activation in NaI detectors for dark matter searches,
  Ph.D. thesis, University of Wisconsin-Madison (2015).

\bibitem{cosmogenic-amare15}
J.~Amare, et~al., Cosmogenic radionuclide production in NaI(Tl) crystals, JCAP
  02 (2015) 046.

\bibitem{cosmogenic-villar18}
P.~Villar, et~al., Study of the cosmogenic activation in NaI(Tl) crystals
  within the ANAIS experiment, Int. J. Mod. Phys. A 33~(9) (2018) 1843006.
  
\bibitem{cosmogenic-amare18}
J.~Amare, et~al., {Cosmogenic production of tritium in dark matter detectors},
  Astropart. Phys. 97 (2018) 96.
    
\bibitem{anais112:2019}
J.~Amare, et~al., {Analysis of backgrounds for the ANAIS-112 dark matter experiment}, 
Eur. Phys. J. C 79 (2019) 412.

\bibitem{gordon:2004}
M.~S.~Gordon, et~al., {Measurement of the Flux and Energy Spectrum of Cosmic-Ray Induced Neutrons
on the Ground}, IEEE Transactions on Nuclear Science 51 (2004) 3427–3434.
M.~S.~Gordon, et~al., Correction to “Measurement of the flux and energy spectrum of cosmic-ray
induced neutrons on the ground”, IEEE Transactions on Nuclear Science 52 (2005) 2703–2703.

\bibitem{Bernabei:2008yh}
R.~Bernabei, et~al., {The DAMA/LIBRA apparatus}, Nucl. Instrum. Meth. A592
  (2008) 297--315.
\newblock \href {http://dx.doi.org/10.1016/j.nima.2008.04.082}
  {\path{doi:10.1016/j.nima.2008.04.082}}.

\bibitem{anais0:2012}
S.~Cebrian, et~al., {Background model for a NaI(Tl) detector devoted to dark
  matter searches}, Astropart. Phys. 37 (2012) 60--69.

\bibitem{anais-modulation:2019}
J.Amare, et~al., {First results on dark matter annual modulation from ANAIS-112
  experiment}, Phys. Rev. Lett. 123, 031301.

\bibitem{cosine100-modulation:2019}
G.~Adhikari, et~al., {(COSINE-100 Collaboration) Search for a dark matter-induced annual modulation signal
  in NaI(Tl) with the COSINE-100 experiment}, Phys. Rev. Lett. 123, 031302.

\end{thebibliography}
